\documentclass[sigconf, nonacm]{acmart}

\usepackage{xspace}
\usepackage{xcolor} 
\usepackage{enumitem}
\usepackage{booktabs}
\usepackage{siunitx}
\usepackage{multirow}
\usepackage{graphicx}
\usepackage{hhline}
\usepackage{makecell}
\usepackage{color, colortbl}
\usepackage{booktabs}
\usepackage{boldline}
\usepackage{pgfplots}
\usepackage{adjustbox} 

\usepgfplotslibrary{groupplots} 
\usepgfplotslibrary{statistics}

\usetikzlibrary{matrix, positioning, pgfplots.groupplots}

\newcommand{\squishlist}{
	\begin{list}{$\bullet$}{
		\setlength{\itemsep}{0pt}
		\setlength{\parsep}{3pt}
		\setlength{\topsep}{3pt}
		\setlength{\partopsep}{0pt}
		\setlength{\leftmargin}{1.2em}
		\setlength{\labelwidth}{1em}
		\setlength{\labelsep}{0.5em}
   }
}

\newcommand{\squishenum}{
	
	\begin{list}{\usecounter{scount}}{
		\setlength{\itemsep}{0pt}
		\setlength{\parsep}{3pt}
		\setlength{\topsep}{3pt}
		\setlength{\partopsep}{0pt}
		\setlength{\leftmargin}{1em}
		\setlength{\labelwidth}{1em}
		\setlength{\labelsep}{0.5em}
	}
}

\newcommand{\squishend}{
	\end{list}
}

\newcommand{\tabincell}[2]{\begin{tabular}{@{}#1@{}}#2\end{tabular}}
\newcolumntype{T}{>{\tiny}c}

\definecolor{g1_c1}{RGB}{230, 85, 13} 
\definecolor{g1_c2}{RGB}{103, 169, 207} 
\definecolor{g1_c3}{RGB}{253, 141, 60}  
\definecolor{g1_c4}{RGB}{76, 199, 128} 
\definecolor{g1_c5}{RGB}{233, 163, 201} 
\definecolor{g1_c6}{RGB}{165, 181, 93} 
\definecolor{g1_c7}{RGB}{179, 132, 186} 
\definecolor{g1_c8}{RGB}{143, 143, 143} 

\definecolor{g2_c1}{RGB}{145, 204, 192} 
\definecolor{g2_c2}{RGB}{127, 171, 209} 
\definecolor{g2_c3}{RGB}{247, 172, 83} 
\definecolor{g2_c4}{RGB}{238, 182, 212} 

\definecolor{g3_c1}{RGB}{143, 143, 143} 
\definecolor{g3_c2}{RGB}{253, 141, 60} 
\definecolor{g3_c8}{RGB}{230, 85, 13} 
\definecolor{g3_c3}{RGB}{103, 169, 207} 
\definecolor{g3_c4}{RGB}{76, 199, 128} 
\definecolor{g3_c5}{RGB}{233, 163, 201} 
\definecolor{g3_c6}{RGB}{165, 181, 93} 
\definecolor{g3_c7}{RGB}{179, 132, 186} 

\definecolor{g4_c1}{HTML}{EA8379} 
\definecolor{g4_c2}{HTML}{7DAEE0} 
\definecolor{g4_c3}{HTML}{B395BD} 

\definecolor{g5_c1}{RGB}{200, 200, 200} 
\definecolor{g5_c2}{HTML}{7ABBDB} 
\colorlet{g5_c3}{g5_c2!60!white} 
\colorlet{g5_c4}{g5_c2!30!white} 

\definecolor{g6_c1}{RGB}{125, 207, 164} 
\definecolor{g6_c2}{RGB}{253, 185, 106} 
\definecolor{g6_c3}{RGB}{143, 143, 143} 
\definecolor{g6_c4}{RGB}{127, 171, 209} 

\definecolor{g7_c1}{RGB}{160, 160, 160} 
\definecolor{g7_c2}{RGB}{253, 141, 60} 
\definecolor{g7_c3}{RGB}{230, 85, 13} 
\definecolor{g7_c4}{RGB}{61, 153, 86} 
\definecolor{g7_c5}{RGB}{78, 178, 238} 

\definecolor{customgreen}{RGB}{77, 175, 74} %
\definecolor{customblue}{RGB}{52, 177, 230} %
\definecolor{custombrown}{RGB}{81, 5, 38} %
\definecolor{customorange}{RGB}{250, 170, 31} %

\newcommand\vldbdoi{XX.XX/XXX.XX}

\newcommand\vldbavailabilityurl{URL_TO_YOUR_ARTIFACTS}

\newcommand\ModelName{\textsf{PRICE}\xspace}
\newcommand\ModelLongName{A Pretrained Model for Cross-Database \\ Cardinality Estimation}

\newcommand{\Card}{{\sf Card}\xspace}
\newcommand{\Dom}{{\sf Dom}\xspace}

\newcommand{\stitle}[1]{\vspace{0.3ex}\noindent\underline{\textbf{#1}}}

\definecolor{mygrey}{RGB}{230,230,240}
\definecolor{myblue}{RGB}{175, 238, 235}

\begin{document}
\pagestyle{plain}
\pagenumbering{arabic}
\title{\ModelName: \ModelLongName}
\author{
	Tianjing Zeng{$^{1}$}, Junwei Lan{$^{1,2,\#}$}, Jiahong Ma{$^{1,3,\#}$}, Wenqing Wei{$^{1,2}$}, Rong Zhu{$^{1,*}$}, Pengfei Li{$^{1}$}, Bolin Ding{$^{1,*}$}, Defu Lian{$^{2}$}, Zhewei Wei{$^{3}$}, Jingren Zhou{$^{1,*}$}}
\affiliation{%
	\institution{{$^{1}$}Alibaba Group, $^{2}$University of Science and Technology of China,  $^{3}$Renmin University of China}
}
\begin{abstract}

Cardinality estimation (CardEst) is essential for optimizing query execution plans. Recent ML-based CardEst methods achieve high accuracy but face deployment challenges due to high preparation costs and lack of transferability across databases. In this paper, we propose \ModelName, a \underline{PR}etrained mult\underline{I}-table \underline{C}ard\underline{E}st model, which addresses these limitations. \ModelName takes low-level but transferable features w.r.t.~data distributions and query information and elegantly applies self-attention models to learn meta-knowledge to compute cardinality in any database. It is generally applicable to any unseen new database to attain high estimation accuracy, while its preparation cost is as little as the basic one-dimensional histogram-based CardEst methods.
Moreover, \ModelName can be finetuned to further enhance its performance on any specific database.

We pretrained \ModelName using 30 diverse datasets, completing the process in about 5 hours with a resulting model size of only about 40MB. Evaluations show that \ModelName consistently outperforms existing methods, achieving the highest estimation accuracy on several unseen databases and generating faster execution plans with lower overhead. After finetuning with a small volume of database-specific queries, \ModelName could even find plans very close to the optimal ones. Meanwhile, \ModelName is generally applicable to different settings such as data updates, data scaling, and query workload shifts. We have made all of our data and codes publicly available at \url{https://github.com/StCarmen/PRICE}. 

\end{abstract}

\maketitle


\vspace{-3pt}
\begingroup
\renewcommand{\thefootnote}{} 
\footnotetext{
$*$ Corresponding authors, primary contact email address red.zr@alibaba-inc.com \\
$\#$ The two authors contribute equally to this paper, listed in alphabetic order. \\}
\renewcommand{\thefootnote}{\arabic{footnote}} 
\endgroup

\section{Introduction}
\label{sec:intro}

Cardinality estimation (CardEst), which predicts the number of tuples of each SQL query before execution, plays a crucial role in query optimization in DBMS. Accurate CardEst could significantly help to identify highly efficient execution plans, thus enhancing the performance of DBMS. As a result, CardEst methods have been extensively studied in the literature, but now their performance is still far from satisfactory.

\stitle{Background and Challenges.}
Traditional methods, including \break one-dimensional (1-D) histograms~\cite{selinger1979access,poosala1997selectivity} and sampling~\cite{leis2017cardinality,heimel2015self}, are widely applied in commercial and open-source DBMSs~\cite{psql2020, sqlserver2019, mysql2020, mdb2020}. They only require gathering statistical features, i.e., the 1-D histogram for each attribute, and thus, they are friendly to deploy for any new database. However, their reliance on oversimplified models and 
unrealistic assumptions on data distributions often leads to poor estimation quality~\cite{DBLP:journals/pvldb/YangLKWDCAHKS19,DBLP:journals/pvldb/ZhuWHZPQZC21}. To address these limitations, a surge in ML-based methods has been proposed in recent years. They either directly learn to model the joint probability density function (PDF) of data to compute cardinalities (data-driven methods~\cite{tzoumas2011lightweight, DBLP:journals/pvldb/ZhuWHZPQZC21}) or build models mapping featurized queries to their cardinalities using collected query workload (query-driven methods~\cite{DBLP:journals/pvldb/DuttWNKNC19, pvldb:DuttWNC20}) or both (hybrid methods~\cite{,li_alece_2023}). Although existing benchmarks and evaluations~\cite{DBLP:journals/pvldb/HanWWZYTZCQPQZL21, kim_learned_2022} have verified that these ML-based methods could attain high accuracy to find better execution plans, their explicit shortcomings still prevent them from being truly applicable:

1) These methods are all data-specific and not transferable across different databases. Data-driven approaches model the specific underlying data distribution for each dataset. Query-driven approaches typically represent features (e.g.,~the values of attributes or table names) as hard encoded vectors, which can not be resolved in other datasets having different tables and attributes. 

2) The preparation cost of existing ML-based methods is much higher than traditional CardEst methods. For each new database, they are required to collect training data, i.e., the sampled tuples in data-driven methods and the executed query workload in query-driven methods, tune hyper-parameters, train the models, and maintain the models for inference, which is time and space-consuming. 


3) The performance of these ML-based methods is not stable~\cite{DBLP:journals/pvldb/HanWWZYTZCQPQZL21, kim_learned_2022}. Data-driven methods often have prior assumptions on data distributions (e.g., DeepDB~\cite{DBLP:journals/pvldb/HilprechtSKMKB20} assumes attributes are not highly correlated), while query-driven models can not perform well on queries having distinct distributions with the training workload~\cite{DBLP:journals/pvldb/ZhuWHZPQZC21}.
Some methods~\cite{DBLP:journals/pvldb/HanWWZYTZCQPQZL21, kim_learned_2022} can not well adapt to dynamic settings.


We summarize the main properties of the above CardEst methods in Table~\ref{tab:comp_methods}. What we desire is a CardEst method that could be effortlessly deployed on any unseen new database with minimal preparations (akin to traditional methods) while still maintaining high but stable accuracy comparable to ML-based methods.

\begin{table*}[!t]
\centering
\caption{\small Comparison of different paradigms of CardEst methods.}
\label{tab:comp_methods}

\vspace{-1em}
\setlength\tabcolsep{5pt}

\resizebox{\linewidth}{!}{%
\arrayrulecolor{black}
\begin{tabular}{c|c|c|c|c}

\hline



\rowcolor{mygrey}
\sf Category of CardEst Methods & \sf Preparations Required to Apply on a New Database & \sf Preparation Cost & \sf Estimation Accuracy & \sf Easy to Apply on Unseen Database \\ \hline

 Traditional Method (\cite{selinger1979access,poosala1997selectivity,li2016wander},$\cdots$) & Collecting Simple  Statistical Features & \bf Low & Low &  \textcolor{green!50!black}{\textbf{\huge\raisebox{-0.5ex}{$\checkmark$}}} 
\\ \hline

{Query-driven Method (\cite{DBLP:conf/sigmod/ZhaoYHLZ22,sigmod:ParkZM20},$\cdots$)} & Workload Collection and Model Training & Medium & High (Possible) &  \textcolor{red!80!black}{\textbf{\huge\raisebox{-0.4ex}{$\times$}}}  \\ \hline

{Data-driven Method(\cite{DBLP:journals/pvldb/YangKLLDCS20,DBLP:journals/pvldb/HilprechtSKMKB20},$\cdots$)} & {Data Distribution  Model Training} & Medium & High (Possible)  & \textcolor{red!80!black}{\textbf{\huge\raisebox{-0.4ex}{$\times$}}}  \\ \hline

\tabincell{c}{Hybrid Method(\cite{DBLP:conf/cidr/KipfKRLBK19,li_alece_2023},$\cdots$)}& \tabincell{c}{Collecting Simple  Statistical Features, \\ Workload Collection and Model Training} & High & High (Possible) & \textcolor{red!80!black}{\textbf{\huge\raisebox{-0.4ex}{$\times$}}} \\ \hline

\tabincell{c}{\cellcolor{g5_c4} \textbf{ Pretrained Model} \\ \cellcolor{g5_c4} \textbf{(Our Method)}} & \tabincell{c}{Collecting Simple Statistical Features \\ and Finetuning (Optional)} & \bf Low & \bf High &\textcolor{green!50!black}{\textbf{\huge\raisebox{-0.5ex}{$\checkmark$}}} \\ \hline

\end{tabular}
} 

\vspace{-8pt}
\end{table*}

\stitle{Motivations and Our Contributions.}
We note that there has been some exploration into developing general models for cost estimation~\cite{hilprecht_zero-shot_2022, agnihotri_zero-shot_2023, heinrich_zero-shot_2022, hilprecht_one_2022} or single-table CardEst~\cite{lu_pre-training_2021}, but our goal is to resolve the most practical, but also complex, CardEst on multi-table join queries. We draw some inspiration from the success of pretrained models in the NLP domain, such as BERT~\cite{devlin-etal-2019-bert} and ChatGPT~\cite{10.5555/3495724.3495883}. They organize information hierarchically from the basic embedding of words (a.k.a.~tokens) to complex semantic representations to tackle numerous NLP tasks simultaneously. In this paper, we lead the pretrained model into multi-table CardEst and propose \ModelName, a \underline{PR}etrained mult\underline{I}-table \underline{C}ard\underline{E}st model that is universally appliable to any database. 

Similar to NLP pretrained models, our \ModelName also apply 
low-level but transferable features to learn high-level intermediate representations for CardEst. From a statistical view, the cardinality of any SQL query could be obtained by fetching the probability of the range specified by its filtering predicates on the joint PDFs of attributes across multiple tables. Therefore, the key tasks are selecting proper features, designing mechanisms to represent the joint PDFs, and fetching the probability by filtering predicates.

To this end, we leverage simple features in \ModelName, including the value distribution of each attribute (represented as 1-D histogram), the distribution of scaling factors of each join condition $T \bowtie S$ (i.e., how many tuples in $S$ are joined for each value in $T$, also represented as 1-D histogram) and some characteristics of each filtering predicate, as basic tokens in our pretrained model. Notably, the feature construction time of our \ModelName is as low as the traditional 1-D histogram-based methods. Meanwhile, unlike existing CardEst methods which are learned to fit each specific database and/or query workload, the feature representations in \ModelName maintain the same semantics across different databases. 

Based on them, we employ a powerful self-attention mechanism to capture the meta-knowledge for CardEst. It learns to simulate the fundamental process to obtain the cardinality by condensing high-level embeddings for building the joint PDFs and filtering the probability,  making it generally applicable to any database.
Specifically, the features w.r.t.~each join condition are firstly fused together as embeddings to represent the backbone information of the table joining multiple tables together. Then, its outputs are combined with the features of other attributes to produce embeddings reflecting the details of the joint PDFs for filtering the probability. The final cardinality is easily obtained upon such embeddings.


Due to the transferable features and the meta-knowledge captured for CardEst in \ModelName, when it is trained over different databases with diversified joint PDFs and filtering conditions, its parameters could be generally applied to any database with reasonable and stable performance. 
To train \ModelName, we gather a broad and varied collection of 30 datasets from various domains, generate $5 \times 10^{4}$ training queries on each of them, and collect their true cardinality for model training. This collection could serve as a new comprehensive benchmark for CardEst. On a common machine, our pretraining consumes only around 5 hours to result in a model with size of around $40$MB. Moreover, the model could be finetuned over each specific database to improve its accuracy further.

By comprehensive evaluations in actual DBMS (PostgreSQL), we find that \ModelName significantly outperforms other CardEst methods. The end-to-end execution time of the query plans generated by \ModelName is consistently better than the original PostgreSQL and comparable to the advanced ML-based methods. After finetuning, the performance of \ModelName is even close to the optimal plans. Meanwhile, we also find that \ModelName is generally applicable to any new databases and different settings, i.e., data updates, data scaling, and query workload shifts. Meanwhile, its time and space costs are much lower than those of other ML-based methods.  

In summary, our main contributions are listed as follows:

1) We design \ModelName, a pretrained multi-table CardEst model that can be easily deployed on any new database with minimal preparations while preserving stable and high estimation accuracy.

2) We collect a new comprehensive benchmark for CardEst and carefully train \ModelName to be applicable.

3)  We conducted extensive experiments to evaluate the performance of \ModelName in various settings, confirming its effectiveness, generality, and robustness.

\stitle{Organizations.}
Section~\ref{sec:problem} discusses the preliminaries on the CardEst problem and existing methods. 
Section~\ref{sec:overview} outlines our roadmap for the pretrained model for CardEst. 
Section~\ref{sec:ourmodel} elaborates the detailed structures of \ModelName.
Section~\ref{sec:training} introduces the details of data collection and model training. 
Section~\ref{sec:experiments} reports the experimental results.
Section~\ref{sec:conclusion} concludes the paper and proposes future work.

\vspace{-5pt}
\section{Preliminaries}
\label{sec:problem}

In this section, we formalize the CardEst problem and review current CardEst methods. Based on this, we summarize some key findings that inspire our work.

\stitle{CardEst Problem.}
A database $D$ comprises a set of tables $T = \{T_{1},\ldots,T_{N}\}$ where each table $T_i$ contains $n_i$ attributes as $T_i = (A_{i, 1}, \dots , A_{i, n_i})$. In this paper, we assume that each attribute $A_{i, j}$ is either categorical or continuous, where a categorical attribute falls into a finite domain $\Dom(A_{i,j})=\{c^{1}_{i,j},\ldots,c^{d}_{i,j}\}$ and a continuous attribute spans in an interval $\Dom(A_{i,j})= [\min_{i, j}, \max_{i, j}]$. 

Given a SQL query $Q$ on database $D$, we seek to estimate the cardinality $\Card(Q)$---the exact number of tuples by executing $Q$ on $D$---without actual execution. Let $Q$ be the most common select-project-join SQL queries on table subset  $T_{Q} \subseteq T$ with a set of join conditions $J$ and filtering predicates $F$, represented as:
\begin{align}
\label{Eq:sub_query_format}
\texttt{SELECT } \text{COUNT}(*) \texttt{ FROM } T_{Q} \texttt{ WHERE } J \texttt{ AND }  F.     
\end{align}
In this paper, we formalize each join condition in $J$ as $T_i.A_{i, x} = T_j.A_{j, y}$, which could either be a primary-foreign key (PK-FK) join or a foreign-foreign key (FK-FK) join. We represent each filtering predicate as 
 ``$T_i.A_{i, j} \textsc{ op value} $'', where $\textsc{op} \in \{<,\geq,>,\leq,=\}$ is a comparison operator and $\textsc{value}$ is picked from $\Dom(A_{i,j})$. 
We reserve the support for other types of join, e.g., non-equal and non-inner join, and $\mathtt{LIKE}$ queries on string attributes for future work, since they are not the main focus of the CardEst problem~\cite{DBLP:journals/pvldb/HanWWZYTZCQPQZL21,li_alece_2023}.


\stitle{Analysis of Existing Methods.}
We summarize existing CardEst methods developed under different technical paradigms in Table~\ref{tab:comp_methods}. The details are reviewed as follows:

\emph{Traditional Methods}, including histogram~\cite{selinger1979access,poosala1997selectivity, deshpande2001independence, gunopulos2000approximating, gunopulos2005selectivity, muralikrishna1988equi, DBLP:journals/pvldb/WuJAPLQR18} and sampling~\cite{leis2017cardinality,heimel2015self,DBLP:journals/pvldb/KieferHBM17,zhao2018random, li2016wander}, are widely applied in commercial and open-source DBMSs~\cite{mysql2020,postgresql,oracle2019,MariaDB2020Statistics}.
They are based on simplified assumptions and expert-designed heuristics and only require gathering very simple statistical features, i.e., the 1-D histogram for each attribute. Hence, these methods are friendly to deploy for any new database, but their reliance on simplified models often leads to poor estimation quality~\cite{DBLP:journals/pvldb/YangLKWDCAHKS19,DBLP:journals/pvldb/ZhuWHZPQZC21}.

\textit{Query-driven Methods} attempt to learn direct mappings from featurized queries to their cardinalities. These methods require the collection of training queries and their corresponding true cardinalities, which usually take a long time. 
Classic methods apply queries to correct and tune histograms~\cite{DBLP:conf/sigmod/BrunoCG01,DBLP:journals/isci/FuchsHL07,DBLP:journals/tkde/KhachatryanMSB15,DBLP:conf/icde/SrivastavaHMKT06} or update statistical summaries~\cite{DBLP:conf/vldb/StillgerLMK01,DBLP:journals/pvldb/WuJAPLQR18}. Recent methods leverage advanced ML models, including DNNs~\cite{DBLP:journals/pvldb/LiuD0Z21,DBLP:conf/sigmod/ZhaoYHLZ22,DBLP:conf/nips/Lakshminarayanan17,DBLP:conf/iclr/LeeBNSPS18, cascon:LiuXYCZ15}, auto-regression~\cite{DBLP:conf/cidr/WuYYZHLLZZ22}, KDE~\cite{DBLP:journals/pvldb/KieferHBM17},  gradient boosted trees~\cite{DBLP:journals/pvldb/DuttWNKNC19, pvldb:DuttWNC20}, etc., to model the complex distributions and improve the estimation accuracy.

\textit{Data-driven Methods} are independent of queries. They learn unsupervised models of the joint distribution of multiple attributes and compute the cardinality using the models. A variety of ML-based models have been used in existing work, including the deep auto-regression model~\cite{DBLP:journals/pvldb/YangLKWDCAHKS19,DBLP:journals/pvldb/YangKLLDCS20,hasan2019multi} and probabilistic graphical models (PGMs) such as
Bayesian networks (BN)~\cite{tzoumas2011lightweight, getoor2001selectivity, DBLP:journals/tit/ChowL68, DBLP:journals/corr/abs-2012-14743}, SPN~\cite{DBLP:journals/pvldb/HilprechtSKMKB20, DBLP:conf/uai/PoonD11}, and FSPN~\cite{DBLP:journals/pvldb/ZhuWHZPQZC21, DBLP:journals/corr/abs-2011-09020}, to model the joint data distribution.

Later, \textit{Hybrid Methods}~\cite{DBLP:conf/sigmod/WuC21, DBLP:conf/cidr/KipfKRLBK19, DBLP:journals/pvldb/NegiWKTMMKA23, DBLP:journals/pvldb/DuttWNKNC19, li_alece_2023} utilize both data distribution as unsupervised information and query workload as supervised information. They exhibit higher estimation accuracy and generality to varying data according to the benchmark evaluations~\cite{DBLP:journals/pvldb/HanWWZYTZCQPQZL21, wang2020ready}.

\stitle{Our Findings and Inspirations.}
As analyzed in Section~\ref{sec:intro}, all these ML-based methods require exhaustive preparation steps (on training data collection and model training) and provide specific models for each new database. They are not friendly for system deployment in production scenarios. Motivated by this, there has been some work dedicated to exploring solutions for cross-database applicability in query optimization, such as the general models for cost estimation~\cite{hilprecht_zero-shot_2022, agnihotri_zero-shot_2023, heinrich_zero-shot_2022, hilprecht_one_2022} or CardEst on a single table~\cite{lu_pre-training_2021}. However, the most intricate and crucial CardEst problem for joining queries on multi-tables remains unsolved. 

To address this challenge, in this paper, we propose a new CardEst method that could be easily applicable to any new, unseen database with minimal or even no preparatory steps. Inspired by the well-established pretraining paradigm in the NLP domain~\cite{devlin-etal-2019-bert,roberta_liu2019roberta,10.5555/3495724.3495883}, 
we embark on an initial foray into developing a pretrained multi-table CardEst model that 
boasts cross-database portability. As shown in Table~\ref{tab:comp_methods}, it could be easily deployed on any unseen database with very little preparations (i.e., gathering some simple statistical information as traditional methods) while still preserving high estimation accuracy comparable to ML-based methods. Meanwhile, we provide the option to finetune this model on each specific database to attain better accuracy. In the following, we present our intuitive roadmap in Section~\ref{sec:overview} and discuss its technical details in Section~\ref{sec:ourmodel}.


\vspace{-5pt}
\section{Our Roadmap}
\label{sec:overview}

To lay the foundations for our pretrained model, in this section, we first dive deep to analyze the CardEst problem and different methods from a more fundamentally statistical perspective (in Section~\ref{sec:overview-1}). Based on the analysis results, in Section~\ref{sec:overview-2}, we present the key preparations we made to design and implement our pretrained model for the CardEst problem. 

\vspace{-5pt}
\subsection{CardEst $\!\!$ Problem $\!\!$ and $\!\!$ Methods: $\!\!$ A Deep Dive}
\label{sec:overview-1}

\stitle{CardEst Problem: A Statistical View.}
For each query $Q$ in the form of Eq.~\eqref{Eq:sub_query_format}, in no ambiguity, we also use $T_{Q}$ to denote the table joining all tables accessed by $Q$. Let $A_{Q}$ be the set of all attributes in $T_{Q}$. We could regard each attribute $A_{i,j} \in A_{Q}$ as a random variable defined over its domain $\Dom(A_{i,j})$. Consequently, the table $T_{Q}$ forms a joint probability density function (PDF) $\Pr(T_{Q}) = \Pr(A_{1,1}, A_{1,2}, \dots, A_{i,j}, \dots)$ on the domain $\Dom(A_{Q}) = A_{1, 1} \times A_{1, 2} \dots \break \times A_{i,j} \times \cdots $. Each record $t \in T_{Q}$ is considered an independent sample drawn from $\Dom(A_{Q})$ by $\Pr(T_{Q})$.
Based on $\Pr(T_{Q})$, the query $Q$ can also be represented in a canonical form as
\begin{align}
\label{Eq:Q}
Q = \{A_{1,1} \in R_{1,1} \wedge A_{1,2} \in R_{1,2} \wedge \cdots \wedge A_{i,j} \in R_{i,j} \wedge \dots \},
\end{align}
where $R_{i,j} \subseteq \Dom(A_{i,j})$ defines the constraint region specified by the filter predicate on attribute $A_{i,j}$. In general, we have $R_{i,j} =  \Dom(A_{i,j})$ if $Q$ has no constraint on $A_{i,j}$. Then, the probability of a 
randomly picked record $t \in T_{Q}$ satisfying query $Q$ is:
\begin{align}
 \Pr(Q) = \Pr(A_{1,1} \in R_{1,1}, A_{1,2} \in R_{1,2}, \dots, A_{i,j} \in R_{i,j}, \dots ).  
\end{align}
Obviously, when $\lvert  T_{Q} \lvert$ hosts a sufficient number of tuples, we have 
\begin{align}
\label{eq:card}
    \Card(Q) = \Pr(Q) \cdot \lvert  T_{Q} \lvert.
\end{align}

\stitle{Traditional CardEst Methods: A Deep Analysis.}
Based on the above statistical model of the CardEst problem, we could dive deep to analyze which information the existing CardEst models exactly learn. 
For data-driven methods, they directly compute $\Pr(Q)$ from $\Pr(T_{Q})$ to obtain $\Card(Q)$ by Eq.~\eqref{eq:card}. 
As $\Pr(T_Q)$ differs for different $Q$, they often apply two approaches to model such joint PDF. 
One way is to model the universal $\Pr(T)$ over all tables in the database $D$~\cite{DBLP:journals/pvldb/YangKLLDCS20}. Certainly, $\Pr(T_{Q})$ is a marginal PDF of $\Pr(T)$ by excluding all attributes $A_{i', j'}$ in tables not occurring in $Q$, so it shrinks $\Pr(T)$ to $\Pr(T_Q)$ and obtains $\Pr(Q)$.
Another way is to model multiple $\Pr(T')$ on a small subset of tables in $D$ and combine them together to obtain $\Pr(T_Q)$~\cite{DBLP:journals/pvldb/HilprechtSKMKB20, DBLP:journals/pvldb/ZhuWHZPQZC21, DBLP:journals/corr/abs-2011-09020}. The key techniques for such models include:
1) how to design a proper model for the joint PDF $\Pr(T)$ (or $\Pr(T')$) such that $\Pr(Q)$ with any range constraints could be efficiently computed;
and 2) how to correct the PDF from $\Pr(T)$ (or $\Pr(T')$) to $\Pr(T_Q)$, i.e., a tuple in $T_Q$ may occur different times in $T$ as it could join with records in other tables (see details in~\cite{DBLP:journals/pvldb/ZhuWHZPQZC21}). 
However, no matter how accurate the learned model for $\Pr(Q)$ is, it only fits for the very specific database $D$.

For query-driven methods, they directly learn the mapping function $\mathcal{F}: Q \to \Card(Q)$. The model $\mathcal{F}$ actually condenses
two types of information as learned parameters together:
1) the joint PDF $\Pr(T_{Q})$; 
and 2) the method to compute $\Pr(Q)$ from $\Pr(T_{Q})$.
The query-driven methods often only use features on tables, joins, and predicates, which only relate to the semantics of $Q$, so the information of $\Pr(T_{Q})$ is learned indirectly. 
This also implies that the feature vectors, as well as the learned parameters, are bound to the database and the training workload, making this learning paradigm unsuitable for other datasets.

The hybrid methods further equip the function $\mathcal{F}$ with additional features on attribute distributions, e.g., the histogram of $\Pr(A_{i, j})$ for each attribute in~\cite{li_alece_2023}, so that it could better capture the data distribution information of $\Pr(T_{Q})$ and exhibit better results than purely query-driven methods~\cite{DBLP:conf/sigmod/ZhaoYHLZ22}.
This paradigm inspires us to believe that the high-dimensional PDF can be learned from the low-dimensional features. However, the query representations and learning paradigm in existing methods are not transferable (i.e.,~keeping the same semantic) across different databases, hindering their potentiality from becoming pretrained models.

\vspace{-5pt}
\subsection{Preparations for the Pretrained Model}
\label{sec:overview-2}

\stitle{Feasibility of the Pretrained Model.}
The above CardEst methods are all data-specific, where both the parameters to construct the joint PDF $\Pr(T_{Q})$ or compute $\Pr(Q)$ are all learned to fit a particular database. However, for a pretrained CardEst model that is applicable to any database, more generalized knowledge must be acquired. To facilitate our model design, we draw some experiences from the NLP domain, where the pretrained models, such as Bert~\cite{devlin-etal-2019-bert} and Chat-GPT~\cite{10.5555/3495724.3495883}, are developed for a long time. In such NLP models, the information is thought to be organized as a hierarchical structure~\cite{devlin-etal-2019-bert, 10.5555/3495724.3495883}. Specifically, the inputs are word embeddings, a.k.a.~tokens, in low-dimensional space. Then, the models are trained on a large corpus to accumulate experience (instantiated as learned parameters) to condense the information of multiple input tokens together to capture the semantics of phrases, sentences, and chapters in high-dimensional space. Such high-level representations could then be easily applied to downstream complex tasks such as text summary or language translation. 

We find that our CardEst problem resembles these NLP tasks: the key information of the joint PDF $\Pr(T_{Q})$ is in high-dimensional space (like sentences); however, it could also be gradually decomposed as the combination of $\Pr(T_i)$ on each table (like phrases), and then $\Pr(A_{i, j})$ on each attribute (like words). The attributes are correlated, and the tables could also be connected through join relations, just as the semantic relations between words, phrases, and sentences. Motivated by this, in our pretrained model for CardEst, we also utilize attribute-level features in low-dimensional space as input features. Then, we apply learned parameters in the models to capture the correlations between attributes and tables to represent the high-level joint PDF $\Pr(T_{Q})$ and compute the probability $\Pr(Q)$ from $\Pr(T_{Q})$. Upon this intuitive roadmap, in the following content, we further analyze the detailed information to be captured in our pretrained model and the proper tools for building our model.

\stitle{Information Captured by the Pretrained Model.}
As we stated above, the pretrained model aims at learning to obtain the probability $\Pr(Q)$ from the high-dimensional joint PDF $\Pr(T_{Q})$ by utilizing a number of features in low-dimensional space w.r.t.~each attribute $A_{i, j} \in A_{Q}$. To this end, the model needs to essentially capture three aspects of key information as follows:

1) \emph{Correlation factors among attributes.}
In a database, attributes in the same or different tables are often correlated with others~\cite{DBLP:journals/pvldb/ZhuWHZPQZC21,DBLP:journals/pvldb/HanWWZYTZCQPQZL21,li_alece_2023}, so we could not simply obtain $\Pr(T_{Q})$ as $\prod_{A_{i, j} \in A_{Q}} \Pr(A_{i, j})$. Instead, we must learn some correction factors to elegantly combine $\Pr(A_{i, j})$ on each singleton attribute together to approximate the joint PDF $\Pr(T_{Q})$, e.g., using addition or multiplication with learned weights to fit the dependency among attributes.

2) \emph{Scaling factors among tables.}
For different tables in a database, the join operation could also change the joint PDF of attributes. As we explained earlier, a tuple in a table $T_i$ may join with a different number of tuples in another table $T_j$. As a result, for each value $v$ of any attribute $A_{i, j}$ in $T_i$, the occurring probability of $v$ in $T_i$ may be different from that in $T_i \bowtie T_j$. Therefore, to combine $\Pr(A_{i, j})$ together as $\Pr(T_{Q})$, we must also consider the correction effects of the join scaling in our learned weights. 

3) \emph{Filtering factors of predicates.}
Upon the joint PDF $\Pr(T_{Q})$, we need to extract the probability $\Pr(Q)$ by the predicates in $Q$. Thus, except simply combining $\Pr(A_{i, j})$ together in terms of the correlation and scaling effects, by Eq.~\eqref{Eq:Q}, the learned weights in the model should also be able to filter $\Pr(A_{i, j})$ to $\Pr(A_{i, j} \in R_{i, j})$, where $R_{i, j}$ is the constraint region specified by the filter predicate on $A_{i, j}$.

Notably, the three key factors (namely correlation, scaling, and filtering) interact with each other, so the correction weights (or model parameters) must be learned altogether to condense their information at the same time. Meanwhile, we require the learned weight parameters to be universal enough. That is, they should be adaptable to attributes and joins with varying levels of correlation and scaling effects, and be applicable to datasets and queries with different numbers of attributes, joins, and filtering predicates. 


\stitle{Tools to Build the Pretrained Model.}
Although the above task seems to be very difficult, we borrow the successful experience of applying the \emph{self-attention mechanism}~\cite{DBLP:conf/nips/VaswaniSPUJGKP17} in the pretrained NLP models~\cite{devlin-etal-2019-bert,roberta_liu2019roberta,10.5555/3495724.3495883} to fulfill such goals. The self-attention mechanism allows models to learn to concentrate on relevant aspects of input data while disregarding the irrelevant, thereby mimicking human cognitive attention processes.

In particular, let $\boldsymbol{X} = [\boldsymbol{x}_1, \boldsymbol{x}_2, \dots, \boldsymbol{x}_n]^\top \in \mathbb{R}^{n \times d}$ be input data with $n$ rows, where each row $\boldsymbol{x}_i$ represents an input token (e.g., a word embedding) with the same dimension $d$. The self-attention mechanism transforms this input matrix into three matrices: Key ($\boldsymbol{K}$), Query ($\boldsymbol{Q}$), and Value ($\boldsymbol{V}$) through learned weight matrices: $\boldsymbol{K} = \boldsymbol{XW}^K$, $\boldsymbol{Q} = \boldsymbol{XW}^Q$ and $\boldsymbol{V} = \boldsymbol{XW}^V$, where $\boldsymbol{W}^K \in \mathbb{R}^{d\times d_k}$, $\boldsymbol{W}^Q \in \mathbb{R}^{d\times d_q}$, and $\boldsymbol{W}^V \in \mathbb{R}^{d\times d_v}$ are learnable matrices for keys, queries, and values, respectively, and $d_q = d_k$ usually.
Denote $\boldsymbol{k}_i$, $\boldsymbol{q}_i$, and $\boldsymbol{v}_i$ as the $i$-th row of the matrices $\boldsymbol{K}$, $\boldsymbol{Q}$, and $\boldsymbol{V}$, respectively. The attention coefficients $\alpha_{i,j}$, which reflect the importance from $\boldsymbol{x}_j$ to $\boldsymbol{x}_i$, are computed as: 
\vspace{-1em}
\begin{align}
	\alpha_{i,j} = \frac{\exp(e_{i,j}/\sqrt{d_k})}{\sum_{l=1}^n \exp(e_{i,l}/\sqrt{d_k})},
\end{align}
where $e_{i,j}$ is the correlation coefficient between query $i$ and key $j$, typically calculated by the dot product of the corresponding vectors $\boldsymbol{q}_i$ and $\boldsymbol{k}_j$: $e_{i,j} = \boldsymbol{q}_i  \boldsymbol{k}_j^\top$ and $\sqrt{d_k}$ is a normalization factor. Based on these scores, we can obtain a number of output tokens $\boldsymbol{y}_1, \boldsymbol{y}_2, \dots, \boldsymbol{y}_n$, where $\boldsymbol{y}_i = \sum_{j} \alpha_{i, j} \boldsymbol{v}_{j}$. Let matrix $\boldsymbol{Y} = {[\boldsymbol{y}_1, \boldsymbol{y}_2, \dots, \boldsymbol{y}_n]}^{\top} \in \mathbb{R}^{n\times d_v}$. The whole self-attention process can be formulated as follows:
\begin{align}
\boldsymbol{Y} = \text{softmax}\left(\frac{\boldsymbol{Q}\boldsymbol{K}^\top}{\sqrt{d_k}}\right) \boldsymbol{V}.
\end{align}
In practice, the self-attention mechanism can be implemented multiple times to create a \emph{multi-head self-attention mechanism}. Specifically, in the $i$-th head, let $\boldsymbol{W}^{K_i}$, $\boldsymbol{W}^{Q_i}$ and $\boldsymbol{W}^{V_i}$ be the key, query and value matrices, respectively, and $\boldsymbol{Y}_i$ be the output matrix. We have
\begin{align}
\boldsymbol{Y}_i = \text{softmax} \left( \frac{({\boldsymbol{XW}^{Q_i})  (\boldsymbol{XW}^{K_i})^\top}}{\sqrt{d_k}} \right) \boldsymbol{XW}^{V_i}.
\end{align}
Then, the final output $\boldsymbol{Y}$ of $H$-head self-attention can be formulated by concatenating the results of all heads with an additional learnable projection matrix $\boldsymbol{W}^Y \in \mathbb{R}^{(Hd_v)\times d}$:
\begin{align}
\boldsymbol{Y} = [\boldsymbol{Y}_1||\boldsymbol{Y}_2|| \dots || \boldsymbol{Y}_{H-1} || \boldsymbol{Y}_H] \boldsymbol{W}^Y.
\end{align}
The multiple heads capture correlations from different perspectives of the input data, allowing the model to flexibly extract information and gain a comprehensive and nuanced understanding of the contexts. Obviously, the (multi-head) self-attention mechanism has three advantages.

First, it could condense the information of all input vectors to the outputs. In the NLP domain, this is applied to compress the semantics of contextual words, phrases, and sentences. For our CardEst problem, it could be applied to:
1) combine correlated low-dimensional information (e.g.,~the value distribution of each attribute and the scaling factor distribution of each join relation) into high-dimensional joint PDF (e.g., ~$\Pr(T_{Q})$;
and 2) combine the joint PDF (e.g.,~$\Pr(T_{Q})$) with filtering conditions to compute the probability $\Pr(Q)$ for CardEst.

Second, for different input embedding vectors, the attention coefficients $\alpha_{i, j}$ are different. This allows the outputs to adaptively focus on different parts of the inputs. In the NLP task, this is useful to resolve the semantics of demonstrative pronouns (e.g.,~it). For our CardEst problem, this mechanism enables us to:
1) compress different distributions of attributes and scaling factors in different ways (e.g.,~process the highly and weakly correlated attributes in different ways);
and 2) extract the probability $\Pr(Q)$ from $\Pr(T_{Q})$  with different weights for different filtering conditions in query $Q$.
This way, the learned model could be generally applicable to different settings.

Third, the size of the learned matrices $\boldsymbol{{W}}^{K}$, $\boldsymbol{{W}}^{Q}$ and $\boldsymbol{{W}}^{V}$ are only determined by the embedding dimensions $d, d_k, d_q, d_v$, but has no relation with the input length $n$. Therefore, it naturally adapts to processing input data with varied lengths without padding or truncation.

Thus, the self-attention module fully attains the requirements to capture the information in the pretrained model, so we apply it as the building blocks in our model design.

\begin{figure*} 
\includegraphics[width=\linewidth]{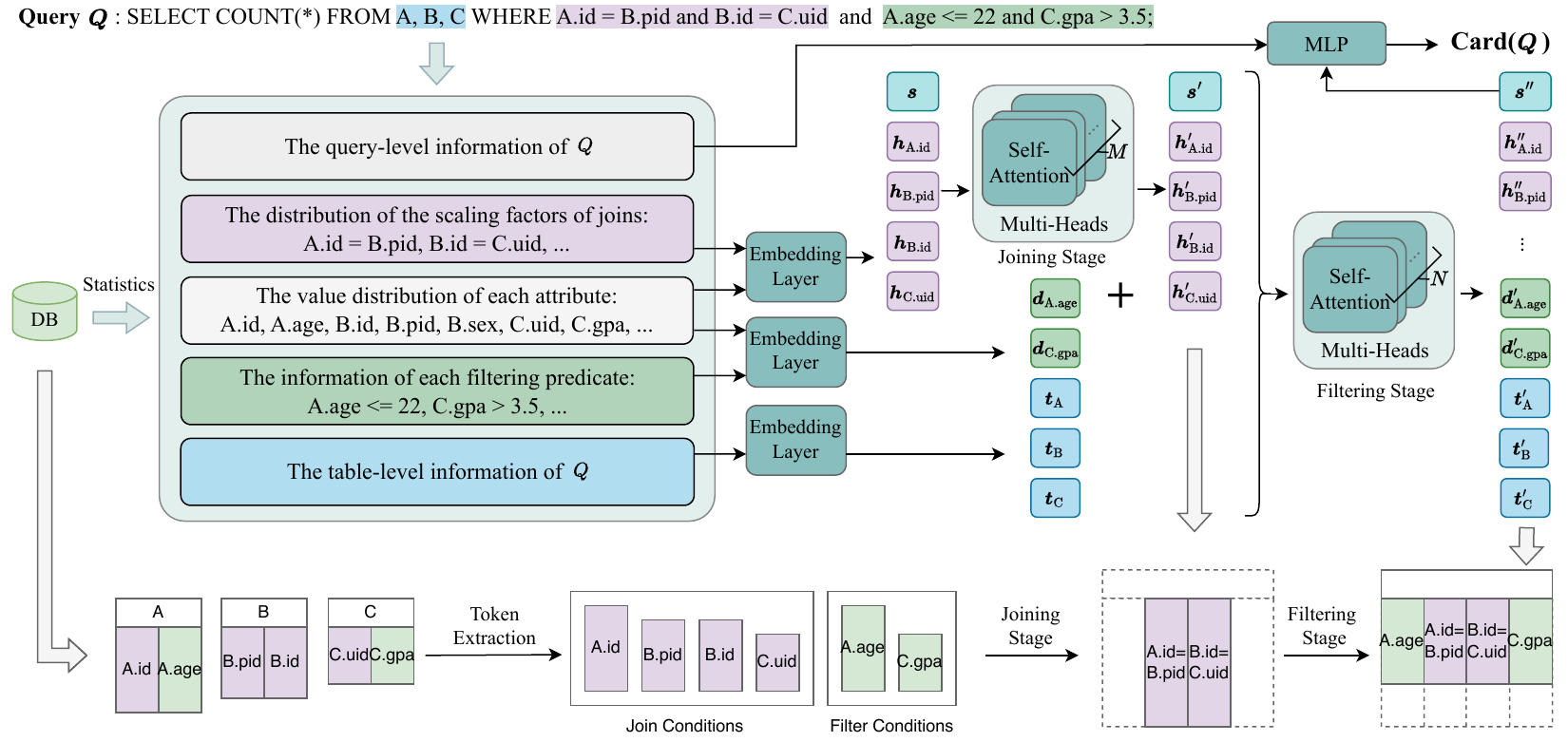}
\vspace{-10pt}
\captionsetup{font=large}
\caption{Model architecture of \ModelName.}
\label{fig:model}
\vspace{-10pt}
\end{figure*} 

\section{Our Pretrained Model}
\label{sec:ourmodel}


We outline the structure and components of our proposed Pretrained Multi-Table CardEst Model (\ModelName) in Figure~\ref{fig:model}. On a high level, for an input query $Q$, \ModelName applies a number of features in low-dimensional space (see details below).
These attributes could be easily obtained at a low cost and widely adapted to any database. 

Then, the model maps such input features into embedding vectors with fixed length and learns to simulate the CardEst process and condense the meta-knowledge for CardEst w.r.t.~the correlation, scaling, and filtering factors as learned parameters. The first \emph{joining stage} simulates the joining process across multiple tables in $Q$ to obtain a backbone structure of the joined table $T_{Q}$ (see Figure~\ref{fig:model}).
It applies a self-attention module to integrate all information w.r.t.~attributes in the join conditions. 
The outputs reflect the effects of joins to the joint PDF $\Pr(T_{Q})$.
The second \emph{filtering stage} utilizes another self-attention module to fuse all relevant information about join conditions and all filtering predicates together to offer a rich representation w.r.t.~$\Pr(T_{Q})$ and $\Pr(Q)$. 
Based on its outputs and some auxiliary query-level features, we could finally obtain the desired cardinality $\Card(Q)$.
The details on feature selection and model workflow are described as follows.

\stitle{Feature Selection.}
For a database $D$ and a SQL query $Q$, our model takes four classes of features  as follows:

1) \emph{The value distribution $\Pr(A_{i, j})$ of each attribute $A_{i, j} \in A_{Q}$.} 
For continuous attributes, we use a histogram vector to represent its distribution. To eliminate the different ranges of attributes, we normalize each histogram vector to span into $[0, 1]$. For categorical attributes, we utilize the SpaceSaving Summary~\cite{10.1145/1166074.1166084} to maintain its distribution. It records the frequency of each item and can be accessed and updated at the same cost as the histogram vector. These vectors of distributions are set to the same length. They are regarded as the basic \emph{tokens} in our approach, similar to the basic word embeddings in the NLP models.

2) \emph{The distribution of the scaling factors of each join condition $T_i.A_{i, x} = T_j.A_{j, y}$ in $Q$.}
Let $s$ be a tuple in table $T_i$. The \emph{scaling factor} of $s$ refers to how many of tuples $t$ in table $T_j$ could join with $s$, i.e.~$s[A_{i, x}] = t[A_{j, y}]$. We also encode the distributions of scaling factors as histogram vectors. They capture the inter-table correlation information and help the model to learn how to scale $\Pr(T_i)$ to $\Pr(T_{Q})$ (see the scaling details in Eq.~(1) in~\cite{DBLP:journals/pvldb/ZhuWHZPQZC21}). 

3) \emph{The information of each filtering predicate} ``$T_i.A_{i, j} \textsc{ op value} $'' \emph{in Q.}
We encode:
i) the range of the filtering \textsc{value} by its lower and upper bound. For categorical attributes, the index of the bin where the \textsc{value} falls in is utilized as the range. If the predicate is an equality, the lower and upper bounds are set to the same value;
and ii) the selectivity, calculated as the ratio of tuples satisfying the predicate on attribute $A_{i, j}$ over all tuples in $T_i$. This aids in computing $\Pr(Q)$ from the joint PDF.

4) \emph{The table-level and query-level auxiliary information of $Q$}.
For each table $T_{i}$ in $Q$, we encode the simplified single-table selectivity estimates produced by heuristic estimators, including AVI, MinSel, and EBO (see details in ~\cite{DBLP:journals/pvldb/DuttWNKNC19}).  We also encode the size $|T_i|$ of each table to characterize its volume.
For query $Q$, we encode:
i) the number of tables and joins in $Q$;
and ii) the cardinality estimated by traditional methods, e.g.,~the histogram-based methods in PostgreSQL.
In this way, we feed the wisdom of the traditional CardEst methods into the model, which could provide more guidelines to produce better results.

Notably, all of the above features could be easily obtained in $O(\sum_{i} |T_i|)$ time. Thus, our \ModelName is as cheap as the efficient traditional 1-D histogram-based methods in terms of feature construction. Meanwhile, when data changes, these features could be incrementally updated in almost real-time.

\stitle{Workflow of \ModelName.}
As shown in Figure~\ref{fig:model}, \ModelName consists of three main stages: the embedding stage for data preparation and the joining and filtering stages for data integration.

\emph{1) Embedding Stage:}
Initially, \ModelName maps input features into embeddings using simple linear models. Specifically, for each attribute $A_{i, j}$ that occurs in any join condition of $Q$, we combine its value distribution vector and the scaling factor distribution vector together as an embedding vector $\boldsymbol{h}_{i, j}$. For each attribute $A_{i, j}$ that occurs in any filtering predicate of $Q$, we combine its value distribution vector, filtering range, and selectivity to output the embedding vector $\boldsymbol{d}_{i,j}$. Notably, for each type of linear model to aggregate information of join attributes or filtering attributes, the parameters are shared across all instances of models. Besides, for the auxiliary information on each table $T_i$, we also apply a linear model with shared parameters to map it into an embedding $\boldsymbol{t}_{i}$. 
Additionally, we randomly initialize a special embedding vector $\boldsymbol{s}$ as input. $\boldsymbol{s}$ would be fed into the subsequent modules with the other embedding vectors together. It serves to amalgamate the information of other embedding vectors together, akin to the CLS token in NLP tasks~\cite{devlin-etal-2019-bert} or the virtual node in Graph Neural Networks (GNN)~\cite{graphormer}. We will explain the detailed role of $\boldsymbol{s}$ later.
Notice that the dimension of all embedding vectors $\boldsymbol{h}_{i,j}$, $\boldsymbol{d}_{i,j}$, $\boldsymbol{s}$ and $\boldsymbol{t}_{i}$ are uniformly fixed, ensuring the generality across different numbers of joins and filtering predicates in the query. Then, \ModelName applies two main stages to exploit the capabilities of the multi-head self-attention mechanism to capture key information for CardEst.

\emph{2) Joining Stage (Backbone Assembly):}
We integrate the information related to all attributes in the join conditions to build the backbone structure of the table $T_{Q}$. The token $\boldsymbol{s}$ and all embedding vectors $\boldsymbol{h}_{i, j}$---representing each attribute $A_{i, j}$ in the join condition---are fed into a multi-head self-attention module. The module converts them into new embedding vectors $\boldsymbol{s}'$ and $\boldsymbol{h}'_{i, j}$. Each $\boldsymbol{h}'_{i, j}$ (as well as $\boldsymbol{s}'$) is a weighted sum over all embedding vectors $\boldsymbol{h}_{i, j}$ and $\boldsymbol{s}$. The weights, a.k.a.~attention coefficients, are determined by learned parameters (see details in Section~\ref{sec:overview-2}). Using this attention mechanism, the output vectors ${\boldsymbol{h}'}_{i, j}$ and $\boldsymbol{s}'$ capture two sides of information together:
i) the correlations between the scaling factor distributions of different joins in $Q$;
and ii) the impact of the joining operations (i.e.,~the scaling effects) on the value distribution of other attributes. 
This combined knowledge is represented as the outputs ${\boldsymbol{h}'}_{i, j}$. Upon them, the model could then know how to scale multiple low-level PDFs $\Pr(T_{i})$ on each single table to the high-level joint PDF $\Pr(T_{Q})$ on the joined table in the subsequent steps.

\emph{3) Filtering Stage (Information Refinement):}
The focus shifts to refining the backbone information with all filtering conditions to obtain accurate estimation results. Specifically, we aim to obtain the information of $\Pr(T_{Q})$ and compute the probability $\Pr(Q)$ from $\Pr(T_{Q})$.
Recall that each embedding vector $\boldsymbol{d}_{i, j}$ condenses the value distribution and filtering information on each attribute $A_{i, j}$ occurring in the filtering predicates, and each embedding vector ${\boldsymbol{h}'}_{i, j}$ condenses the information of the scaling factor distributions of all joins. In this stage, we apply another multi-head self-attention module to integrate their information together.
In particular, let $U =  \{\boldsymbol{d}_{i, j}\} \cup \{ \boldsymbol{h}'_{i, j} \} \cup \{ \boldsymbol{s}' \} \cup \{ \boldsymbol{t}_{i} \}$ be the set of all input embedding vectors. Then, for each $\boldsymbol{u} \in U$, the self-attention module obtains an embedding vector $\boldsymbol{u}'$ using a weighted sum over all vectors in $U$.

\begin{scriptsize}

\begin{table*}[htbp]
\centering

\caption{Overview of datasets: database statistics and workload specifications.}
\label{tab:datasets}

\setlength\tabcolsep{2pt}
\renewcommand{\arraystretch}{1.2}

\resizebox{\textwidth}{!}{%
\arrayrulecolor{black}
\begin{tabular}{c|cccccccccc|ccTc}

\hline
\rowcolor{mygrey}
\cellcolor{mygrey} & \multicolumn{10}{c|}{\cellcolor{mygrey} \sf Database Statistics} & \multicolumn{4}{c}{\cellcolor{mygrey} \sf Workload Specifications} \\ \cline{2-15}
\rowcolor{mygrey} 
\cellcolor{mygrey} & \cellcolor{mygrey} & \cellcolor{mygrey} & \sf \# of Cols & \sf \# of Rows & \sf \# of Join & \cellcolor{mygrey} & \sf Total Attribute & \sf Distribution Skewness & \sf Average Pairwise & \sf Join & \sf Joined & \# \sf of Filtering & \cellcolor{mygrey} & \sf True Cardinality \\
\rowcolor{mygrey} 
\multirow{-3}{*}{\cellcolor{mygrey} \sf Dataset Name} & \multirow{-2}{*}{\cellcolor{mygrey} \sf Sectors} & \multirow{-2}{*}{\cellcolor{mygrey}\sf \# of Tables} & \sf (All Tables) & \sf (All Tables) & \sf Relations & \multirow{-2}{*}{\cellcolor{mygrey} \sf Volume} & \sf Domain Size & \sf (MIN/AVG/MAX) & \sf Correlation & \sf Forms & \sf Tables & \sf Predicates & \multirow{-2}{*}{\cellcolor{mygrey} \sf \scriptsize{Join Type}} & \sf Range \\ 
\hline





 
Accidents & Government & 3 & 43 & $1.5 \cdot 10^{6}$ & 3 & 118.6M & $1.8 \cdot 10^{6}$ & -0.4/1.3/8.1 & 0.33 & Chain & 2 - 3 & 0 - 7 & PK-FK/FK-FK & $10^1$ - $6$$\cdot$$10^{9}$ \\

Airline & Retail & 19 & 119 & $1.1 \cdot 10^{7}$ & 22 & 4.5G & $1.2 \cdot 10^{5}$ & -2.2/5.1/129.8 & 0.17 & Star & 2 - 7 & 2 - 7 & PK-FK & $10^1$ - $1$$\cdot$$10^{7}$ \\

Baseball & Sport & 25 & 359 & $4.7 \cdot 10^{5}$ & 103 & 30.6M & $3.1 \cdot 10^{5}$ &  -13.6/1.7/101.7 & 0.38 & Mixed & 2 - 7 & 0 - 7 & PK-FK/FK-FK & $10^1$ - $6$$\cdot$$10^{9}$ \\

Basketball & Sport & 9 & 198 & $4.5 \cdot 10^{4}$ & 32 & 4.8M & $1.2 \cdot 10^{5}$ & -4.4/2.6/90.0 & 0.53 & Mixed & 2 - 7 & 0 - 6 & PK-FK/FK-FK & $10^1$ - $1$$\cdot$$10^{10}$ \\

Carcinogenesis & Medicine & 6 & 24 & $2.8 \cdot 10^{4}$ & 15 & 776.0K & $5.3 \cdot 10^{4}$ & -0.2/0.0/0.2 & 0.69 & Star & 2 - 6 & 0 - 6 & PK-FK/FK-FK & $10^1$ -  $5$$\cdot$$10^{7}$ \\

CCS & Financial & 5 & 21 & $4.2 \cdot 10^{5}$ & 5 & 8.1M & $3.5 \cdot 10^{5}$ & -1.6/2.2/18.5 & 0.35 & Mixed & 2 - 5 & 0 - 10 & PK-FK/FK-FK & $10^1$ - $4$$\cdot$$10^{5}$ \\

ChEMBL & Medicine & 9 & 80 & $3.7 \cdot 10^{6}$ & 19 & 1.7G & $8.4 \cdot 10^{6}$ & -12.7/93.6/1341.4 & 0.52 & Mixed & 2 - 7 & 0 - 10 & PK-FK/FK-FK & $10^1$ - $2$$\cdot$$10^{9}$ \\

Consumer & Retail & 3 & 25 & $2.2 \cdot 10^{6}$ & 3 & 78.9M & $3.0 \cdot 10^{6}$ & -1.6/5.9/106.4 & 0.40 & Chain & 2 - 3 & 1 - 6 & PK-FK/FK-FK & $10^1$ - $6$$\cdot$$10^{6}$ \\

Credit & Synthetic & 8 & 73 & $1.6 \cdot 10^{6}$ & 13 & 82.0M & $1.8 \cdot 10^{6}$ & -1.8/0.0/2.3 & 0.21 & Mixed & 2 - 7 & 0 - 6 & PK-FK/FK-FK & $10^1$ - $3$$\cdot$$10^{9}$ \\

Employee & Synthetic & 6 & 25 & $3.9 \cdot 10^{6}$ & 11 & 134.1M & $1.3 \cdot 10^{6}$ & -0.2/0.2/0.8 & 0.04 & Mixed & 2 - 6 & 0 - 3 & PK-FK/FK-FK & $10^1$ - $5$$\cdot$$10^{6}$ \\

Financial & Financial & 8 & 55 & $1.1 \cdot 10^{6}$ & 11 & 65.1M & $1.3 \cdot 10^{6}$ & -0.1/1.4/7.6 & 0.37 & Mixed & 2 - 7 & 0 - 7 & PK-FK/FK-FK & $10^1$ - $3$$\cdot$$10^{8}$ \\

FNHK & Medicine & 3 & 24 & $2.1 \cdot 10^{6}$ & 3 & 68.6M & $1.9 \cdot 10^{5}$ & -1.0/8.6/46.2 & 0.24 & Chain & 2 - 3 & 0 - 7 & PK-FK/FK-FK & $10^1$ - $6$$\cdot$$10^{7}$ \\

Grants & Education & 12 & 51 & $3.0 \cdot 10^{6}$ & 21 & 701.0M & $3.6 \cdot 10^{6}$ & -1.2/25.5/340.3 & 0.22 & Mixed & 2 - 7 & 0 - 10 & PK-FK/FK-FK & $10^1$ - $2$$\cdot$$10^{6}$ \\

Hepatitis & Medicine & 7 & 26 & $1.3 \cdot 10^{4}$ & 9 & 224.0K & $1.4 \cdot 10^{4}$ & -0.2/0.1/0.8 & 0.25 & Mixed & 2 - 7 & 0 - 7 & PK-FK/FK-FK & $10^1$ - $1$$\cdot$$10^{4}$ \\

Hockey & Sport & 18 & 300 & $9.5 \cdot 10^{4}$ & 145 & 9.4M & $9.0 \cdot 10^{4}$ & -1.9/0.5/5.6 & 0.33 & Mixed & 2 - 7 & 0 - 9 & PK-FK/FK-FK & $10^1$ - $2$$\cdot$$10^{7}$ \\

LegalActs & Government & 5 & 33 & $1.8 \cdot 10^{6}$ & 8 & 188.8M & $2.0 \cdot 10^{6}$ & -15.3/-0.3/5.4 & 0.22 & Mixed & 2 - 4 & 0 - 10 & PK-FK/FK-FK & $10^1$ - $1$$\cdot$$10^{6}$ \\

MovieLens & Entertainment & 7 & 24 & $1.2 \cdot 10^{6}$ & 9 & 17.8M & $2.1 \cdot 10^{5}$ & -1.7/-0.4/0.1 & 0.12 & Mixed & 2 - 7 & 0 - 7 & PK-FK/FK-FK & $10^1$ - $5$$\cdot$$10^{7}$ \\

Sakila & Synthetic & 15 & 86 & $4.6 \cdot 10^{4}$ & 31 & 2.9M & $1.2 \cdot 10^{5}$ & -6.0/-0.1/0.5 & 0.18 & Mixed & 2 - 7 & 0 - 10 & PK-FK/FK-FK & $10^1$ - $3$$\cdot$$10^{10}$ \\

SAP & Synthetic & 5 & 45 & $4.1 \cdot 10^{6}$ & 6 & 172.7M & $4.6 \cdot 10^{6}$ & -0.8/-0.1/1.0 & 0.54 & Star & 2 - 4 & 0 - 10 & PK-FK/FK-FK & $10^1$ - $3$$\cdot$$10^{6}$ \\

Seznam & Retail & 4 & 14 & $2.7 \cdot 10^{6}$ & 6 & 67.4M & $3.1 \cdot 10^{5}$ & 0.7/22.0/101.1 & 0.19 & Star & 2 - 4 & 0 - 7 & PK-FK/FK-FK & $10^1$ - $3$$\cdot$$10^{8}$ \\

SSB & Synthetic & 5 & 58 & $1.3 \cdot 10^{7}$ & 4 & 1.2G & $1.2 \cdot 10^{7}$ & -5.3/0.1/5.8 & 0.13 & Star & 2 - 5 & 0 - 7 & PK-FK & $10^1$ - $1$$\cdot$$10^{7}$ \\

TalkingData & Technology & 8 & 25 & $6.9 \cdot 10^{7}$ & 10 & 2.1G & $7.5 \cdot 10^{6}$ & -1.4/0.0/1.4 & 0.65 & Mixed & 2 - 7 & 0 - 10 & PK-FK/FK-FK & $10^1$ - $8$$\cdot$$10^{9}$ \\

Telstra & Government & 5 & 12 & $1.5 \cdot 10^{5}$ & 10 & 2.9M & $9.4 \cdot 10^{4}$ & 0.0/1.8/11.6 & 0.05 & Star & 2 - 5 & 0 - 5 & PK-FK/FK-FK & $10^1$ - $2$$\cdot$$10^{5}$ \\

Tournament & Sport & 9 & 106 & $2.0 \cdot 10^{5}$ & 21 & 9.6M & $7.1 \cdot 10^{3}$ & -0.4/0.6/7.2 & 0.17 & Mixed & 2 - 7 & 0 - 7 & PK-FK/FK-FK & $10^1$ - $2$$\cdot$$10^{9}$ \\

TPC-H & Synthetic & 8 & 61 & $8.7 \cdot 10^{6}$ & 12 & 1.1G & $1.5 \cdot 10^{7}$ & 0.0/0.1/0.6  & 0.08 & Mixed & 2 - 7 & 0 - 7 & PK-FK/FK-FK & $10^1$ - $5$$\cdot$$10^{9}$ \\

TubePricing & Education & 6 & 40 & $1.5 \cdot 10^{5}$ & 9 & 5.7M & $1.9 \cdot 10^{5}$ & -1.9/17.5/240.2 & 0.51 & Mixed & 2 - 6 & 0 - 10 & PK-FK/FK-FK & $10^1$ - $1$$\cdot$$10^{5}$ \\

\hline

IMDB & Entertainment & 6 & 37 & $6.2 \cdot 10^{7}$ & 15 & 3.0G & $8.5 \cdot 10^{7}$ & -1.9/19.2/384.5 & 0.32 & Star & 2 - 5 & 0 - 4 & PK-FK/FK-FK & $2$ - $1$$\cdot$$10^{10}$ \\

STATS & Education & 8 & 43 & $1.0 \cdot 10^{6}$ & 30 & 32.3M & $1.9 \cdot 10^{6}$ & -1.0/10.5/134.5 & 0.46 & Mixed & 2 - 7 & 0 - 10 & PK-FK/FK-FK & $16$ - $8$$\cdot$$10^{10}$ \\

ErgastF1 & Sport & 14 & 98 & $5.5 \cdot 10^{5}$ & 67 & 15.1M & $3.2 \cdot 10^{5}$ & -2.0/2.2/66.6 & 0.44 & Mixed & 2 - 7 & 0 - 9 & PK-FK/FK-FK & $1$ - $2$$\cdot$$10^{10}$ \\

VisiualGenome & Education & 6 & 20 & $3.6 \cdot 10^{6}$ & 6 & 73.4M & $3.1 \cdot 10^{5}$ & -3.8/-0.4/2.0 & 0.20 & Mixed & 2 - 6 & 0 - 7 & PK-FK/FK-FK & $24$ - $2$$\cdot$$10^{8}$ \\

\hline

\end{tabular}

} 

\end{table*}

\end{scriptsize}

We find that, the vectors $\boldsymbol{u}'$ integrates three sides of information together:
i) the scaling information of multiple tables (encoded in $\boldsymbol{h}'_{i, j}$) are fused with the value distributions of other attributes (encoded in $\boldsymbol{d}_{i, j}$). This comprehensively amalgamates the information to represent the joint PDF $\Pr(T_{Q})$;
ii) the information of all filtering predicates (also encoded in $\boldsymbol{d}_{i, j}$), which reflects how to compute $\Pr(Q)$ from $\Pr(T_{Q})$;
and iii) some coarsen-grained information for estimated cardinality on each single table (encoded in $\boldsymbol{t}_{i}$). This would more or less provide the representation vectors some guidelines from the traditional CardEst methods to correct $\Pr(Q)$.
In short, the output vectors $\boldsymbol{u}'$ encapsulate all relevant information about join conditions and filtering predicates to offer a rich representation w.r.t.~$\Pr(T_{Q})$ and $\Pr(Q)$. 

Based on these, we are prepared to obtain the final estimated cardinality. Notice that, it is not necessary to utilize all output vectors $\boldsymbol{u}'$ produced by the filtering stage. Instead, let $\boldsymbol{s}''$ be the corresponding output vector to $\boldsymbol{s}'$. We only need to apply $\boldsymbol{s}''$ into the final step to reduce the computation cost. This is because $\boldsymbol{s}''$ already condenses the information of other input vectors through the self-attention mechanism. We could regard it as a high-order summarization of the outputs (as well as the input features).  
As a result, we only feed $\boldsymbol{s}''$ and all auxiliary query-level features (serving as guidelines) into a non-linear neural network, specifically referring to a multilayer perceptron (MLP)~\cite{DBLP:books/lib/HastieTF09}, to obtain the desired cardinality $\Card(Q)$.

\stitle{Advantages of \ModelName.}
According to its model design, \ModelName attains high generality.
First, by applying the self-attention mechanism, the learned parameters adapt to queries with any number of joins, tables, and filtering predicates.
Second, we standardize the features w.r.t.~data as a number of normalized vectors of distributions, which hinder the discrepancies in the magnitude of data size and attribute domain size.
Third, \ModelName learns condense key information w.r.t.~correlations, scaling, and filtering factors for CardEst. Such meta-knowledge is independent of data and query workload.
Thus, as long as the model parameters are well-trained in a variety of databases to witness enough different data distributions and query workloads, the model parameters could be transferable. The preparation cost to collect features to apply the pretrained model is as low as the histogram-based CardEst methods. We discuss the details of data collection and model training in Section~\ref{sec:training}. The evaluation results in Section~\ref{sec:experiments} exhibit that our \ModelName could be applied in dynamic data, changing workload, or even unseen new database.

Further, \ModelName also reserves the flexibility to fit each specific database through finetuning. In this way, its performance could be further improved. 
As verified in Section~\ref{sec:experiments}, on unseen databases, the original \ModelName could already achieve commendable performance, while the finetuned \ModelName could attain state-of-the-art results.

\vspace{-10pt}
\section{Data Collection and Model Training}
\label{sec:training}

In this section, we introduce how to collect different databases and the related query workloads to train our model \ModelName. We have made both the collected databases and query workloads and the pretrained model publicly available in the repository~\cite{price_codes_url} to nourish the community. We hope the collected databases and query workloads could serve as a new and comprehensive benchmark for evaluating CardEst methods.

\stitle{Datasets Collection.}
To effectively assess the performance of our \ModelName, it is imperative to train and test the model across various datasets. However, existing CardEst methods commonly utilize a limited number of datasets for evaluation, which are insufficient for the expansive training needs of the pretrained model. To cultivate a model capable of generalizing across diversified databases, we exhaustively search the public data repositories and obtain 30 datasets, including:
1) 6 well-established benchmark datasets for evaluating CardEst methods, such as TPC-H~\cite{tpch_url}, SSB~\cite{10.1007/978-3-642-10424-4_17}, IMDB~\cite{imdb_url} and STATS~\cite{STATS_url};
and 2) 24 new datasets assembled from~\cite{DBLP:journals/corr/MotlS15}.
We clean and process each dataset. Their detailed properties are listed in Table~\ref{tab:datasets}. We observe that these datasets closely resemble the real-world scenarios and demonstrate notable diversity, including:

1) \emph{Domain}: These datasets include both synthetic and real-world data. For real-world data, it spans various domains such as government, sports, retail, and medicine, each differing significantly in scale, data distribution, and schema complexity (see details below). 

2) \textit{Scale}: The collection includes datasets in different volumes (from 2.9MB to 4.5GB); different numbers of tables (from 3 to 25), columns (from 12 to 359), and rows(from $1.29 \times 10^{4}$ to $6.85 \times 10^{7}$); and different domain size of attributes (from $7.1 \times 10^3$ to $1.44 \times 10^{7}$).

3) \textit{Data Distribution}: 
The datasets exhibit significantly different levels of correlations among distributions of attributes. The average skewness scores of attribute distributions span from 0 to 93.6, and the average pairwise attribute RDC scores~\cite{lopez-paz_randomized_nodate} (measuring the correlation) span from 0.04 to 0.53. This allows the pretrained model to learn different patterns to fit different joint PDFs.

4) \textit{Join Scheme}: 
All datasets (except Airline and SSB) contain both primary-foreign key (PK-FK) and foreign-foreign key (FK-FK) joins. The structures of the join schema include the simple chain and star form of tables, as well as the complex forms in a mixture of chains and stars of tables. 

\stitle{Query Workloads Generation.}
For model training and evaluation, corresponding workloads for each dataset must be prepared.
We retain the testing workloads JOB-light\cite{job_light_url} and STATS-CEB\cite{DBLP:journals/pvldb/HanWWZYTZCQPQZL21} for widely recognized datasets IMDB and STATS, respectively, and developed new training and testing workloads for others by the following process:

Specifically, for each dataset, we obtain a graph of its join schema where each node represents a table $T_i$ and each edge connecting $T_i$ and $T_j$ represents a join relation between two tables. We enumerate all connected subgraphs from the join schema graph, where each subgraph represents a possible join relation among some tables in the dataset. To generate an SQL query, we randomly select a subgraph and then attach it with some filtering predicates. In particular, we first collect all categorical and numerical attributes in all the tables that occurred in the subgraph. Let $m$ be the total number of attributes. We uniformly sample a number $n$ from $1$ to $m$ and then repeatedly sample $n$ attributes at random. For each sampled attribute $A_i$, if it is numerical, we pick its lower bound $l$ and upper bound $u$ uniformly at random and set the filtering predicate as $A_i \leq u \text{ and } A_i \geq l$; if it is categorical, we randomly select a value $v$ from its domain and set the filtering predicate as $A_i = v$. For each dataset, we generate $5 \times 10^{4}$ SQL queries for model training. We obtain the true cardinality of each query through actual execution.

\stitle{Model Pretraining.}
We reserve the four widely applied datasets---IMDB, STATS, ErgastF1 and VisualGenome---as the test datasets and apply the remaining 26 datasets with the generated query workloads to pretrain our model \ModelName. The details are as follows:

1) \emph{Loss Function}: 
Due to the board range of the cardinality (e.g., from $1$ to $10^{11}$), we apply a logarithmic transformation to normalize its scale and use $\log{(\Card(Q))}$ as the target label for training. The Mean Squared Error (MSE) function is employed as the loss function during training. That is, for a batch of $k$ SQL queries $Q_1, Q_2, \dots, Q_k$, we have 
\begin{equation}
    \mathsf{MSE\_Loss} = \frac{1}{k} \sum_{k'=1}^{k} {(\log{(\Card(Q_{k'}))} - \log{(\widehat{\Card}(Q_{k'}))})}^{2}.
\end{equation}

2) \emph{Model Hyperparameters}:
We implement our model and its training function in Python 3.10.
In our \ModelName, we use a 40-dimensional vector to represent all input features on value distribution and scaling factor distribution of attributes. The dimension of all inner embedding vectors $\boldsymbol{h}_{i, j}$, ${\boldsymbol{h}'}_{i, j}$, $\boldsymbol{d}_{i, j}$, $\boldsymbol{s}$, $\boldsymbol{s}'$, $\boldsymbol{s}''$ and $\boldsymbol{t}_{i}$ (see details in Section~\ref{sec:ourmodel}) are set to 256. The self-attention modules applied in the joining and filtering stage all comprise 8 heads.

3) \emph{Training Method}:
To promote stable learning and robust generalization across databases, we shuffle the training data and use large batch size (i.e.,~$1.5 \times 10^{4}$). We adopt the Adam optimizer~\cite{DBLP:journals/corr/KingmaB14} for model training with the initial learning rate of $2.85 \times 10^{-5}$ and a weight decay of $5 \times 10^{-5}$. To stabilize the training process, we utilize the StepLR function in PyTorch to adaptively adjust the learning rate. To prevent overfitting, we also apply the dropout function.

4) \emph{Training Environment}:
We train our model on a Linux server outfitted with a 96-core Xeon(R) Platinum 8163 CPU @ 2.50GHz, 512GB of main memory, and 8 NVIDIA A100 GPUs.
For the whole training workload having $1.3 \times 10^{6}$ SQL queries, the pretraining stage only consumes around 5 hours. The pretrained model is only around $40$MB.

\vspace{-5pt}

\section{Experimental Studies}
\label{sec:experiments}

This section presents a comprehensive evaluation of \ModelName against existing state-of-the-art (SOTA) techniques for multi-table cardinality estimation. The evaluation results aim to answer the following pivotal questions about its performance:
 
1) How does our pretrained model \ModelName perform when directly applied to unseen databases with little preparation? (Section~\ref{sec:exp-pretrain})

2) How does \ModelName perform after finetuning on specific databases compared to existing ML-based CardEst methods?  (Section~\ref{sec:exp-finetune})

3) Can \ModelName generalize to data updates and query workload drifts?  (Section~\ref{sec:exp-robust})

4) What are the impacts of training data to \ModelName?  (Section~\ref{sec:exp-dataset})


\vspace{-5pt}
\subsection{Experiment Setup}
\label{sec:exp-setup}



\begin{scriptsize}
\begin{table*}[htbp]
\centering
\caption{Overall performance of different CardEst methods.}
\label{fig:pretrain}
\vspace{-8pt}
\setlength\tabcolsep{2.5pt}

\resizebox{\textwidth}{!}{%

\arrayrulecolor{black}

\begin{tabular}{
  c | c | c |
  c c c c c | c c c c c |
  c | c | c
  }

\hline

\rowcolor{mygrey}
\cellcolor{mygrey}& \cellcolor{mygrey}& \cellcolor{mygrey}& \multicolumn{5}{c|}{\cellcolor{mygrey}\sf Q-ERROR} & \multicolumn{5}{c|}{\cellcolor{mygrey}\sf P-ERROR} & \cellcolor{mygrey}& \cellcolor{mygrey}& \cellcolor{mygrey}\\ \cline{4-13}
\rowcolor{mygrey}
\multirow{-2}{*}{\cellcolor{mygrey} \sf DATASETS} & \multirow{-2}{*}{\cellcolor{mygrey} \sf CARDEST METHOD} & \multirow{-2}{*}{\cellcolor{mygrey}\begin{tabular}[c]{@{}c@{}} \sf E2E\\ \sf TIME (S)\end{tabular}} & 
\sf 50\% & \sf 80\% & \sf 90\% & \sf 95\% & \sf 99\% & \sf 50\% & \sf 80\% & \sf 90\% & \sf 95\% & \sf 99\% & \multirow{-2}{*}{\cellcolor{mygrey}\begin{tabular}[c]{@{}c@{}}\sf MODEL\\ \sf SIZE (MB)\end{tabular}} & \multirow{-2}{*}{\cellcolor{mygrey}\begin{tabular}[c]{@{}c@{}} \sf (PRE)TRAINING\\ \sf TIME (MIN)\end{tabular}} & \multirow{-2}{*}{\cellcolor{mygrey}\begin{tabular}[c]{@{}c@{}}\sf INFERENCE\\ \sf TIME (MS)\end{tabular}} \\

\hline

\multirow{9}{*}{IMDB}
  & {PG} & \cellcolor{g5_c1}$4037$ & $1.95$ & $6.04$ & $19.04$ & $49.44$ & $879.64$ & $\mathbf{1.00}$ & $1.31$ & $1.73$ & $2.45$ & $13.44$ & $-$ & $-$ & $-$ \\

  & {ALECE} & $3911$ & $1.75$ & $4.21$ & $11.49$ & $19.07$ & $124.68$ & $\mathbf{1.00}$ & $1.07$ & $1.46$ & $1.92$ & $5.44$ & $233.11$ & $472.88$ & $5.08$ \\

  & {MSCN} & $4755$ & $3.24$ & $12.75$ & $28.54$ & $104.20$ & $411.90$ & $1.07$ & $1.71$ & $2.31$ & $4.00$ & $15.66$ & $\mathbf{1.57}$ & $75.19$ & $\mathbf{0.79}$ \\

  & {DeepDB} & $3850$ & $1.31$ & $2.97$ & $3.61$ & $\mathbf{5.03}$ & $\mathbf{14.03}$ & $\mathbf{1.00}$ & $\mathbf{1.00}$ & $1.09$ & $1.27$ & $1.56$ & $88.33$ & $73.73$ & $4.73$ \\

  & {NeuroCard} & $3664$ & $1.66$ & $4.14$ & $7.80$ & $14.25$ & $22.22$ & $\mathbf{1.00}$ & $1.23$ & $1.78$ & $2.32$ & $4.05$ & $50.27$ & $14.55$ & $18.43$ \\

  & {FactorJoin} & \cellcolor{g5_c4}$3588$ & $13.86$ & $89.88$ & $348.96$ & $1027.19$ & $4794.80$ & $1.04$ & $1.61$ & $2.86$ & $3.46$ & $7.91$ & $5.59$ & $78.45$ & $3.20$ \\

  & {\ModelName (Pretrained)} & $3910$ & $1.77$ & $4.07$ & $8.39$ & $15.45$ & $70.88$ & $\mathbf{1.00}$ & $1.05$ & $1.16$ & $1.28$ & $1.63$ & $41.14$ & $313.56$ & $5.27$ \\

  & {\ModelName (Finetuned)} & \cellcolor{g5_c3}$\mathbf{3454}$ & $\mathbf{1.29}$ & $\mathbf{2.05}$ & $\mathbf{2.92}$ & $5.36$ & $29.45$ & $\mathbf{1.00}$ & $\mathbf{1.00}$ & $\mathbf{1.04}$ & $\mathbf{1.08}$ & $\mathbf{1.16}$ & $41.14$ & $\mathbf{6.47}$ & $5.27$ \\

  & {Optimal} & \cellcolor{g5_c2}$3442$ & $1.00$ & $1.00$ & $1.00$ & $1.00$ & $1.00$ & $1.00$ & $1.00$ & $1.00$ & $1.00$ & $1.00$ & $-$ & $-$ & $-$ \\
\hline

\multirow{9}{*}{STATS}
  & {PG} & \cellcolor{g5_c1}$30484$ & $1.87$ & $7.11$ & $20.71$ & $73.35$ & $1600.22$ & $1.04$ & $1.67$ & $2.44$ & $4.16$ & $20.57$ & $-$ & $-$ & $-$ \\

  & {ALECE} & $25716$ & $1.67$ & $3.79$ & $7.93$ & $16.44$ & $118.96$ & $\mathbf{1.00}$ & $1.38$ & $1.84$ & $2.17$ & $3.49$ & $233.04$ & $93.09$ & $10.88$ \\

  & {MSCN} & $42051$ & $6.19$ & $70.02$ & $363.71$ & $1609.45$ & $9.15$$\cdot$$10^4$ & $1.36$ & $2.64$ & $5.98$ & $13.41$ & $59.14$ & $\mathbf{1.60}$ & $2.03$ & $\mathbf{0.81}$ \\

  & {DeepDB} & $22931$ & $1.84$ & $19.94$ & $73.52$ & $132.63$ & $1507.78$ & $1.03$ & $2.17$ & $2.85$ & $4.41$ & $9.32$ & $249.93$ & $142.12$ & $12.42$ \\

  & {NeuroCard} & $24435$ & $2.12$ & $12.57$ & $48.40$ & $228.41$ & $1.32$$\cdot$$10^4$ & $1.03$ & $1.56$ & $1.97$ & $2.92$ & $8.22$ & $121.88$ & $36.43$ & $33.80$ \\

  & {FactorJoin} & $50179$ & $5.41$ & $40.44$ & $150.52$ & $666.59$ & $2.10$$\cdot$$10^4$ & $1.09$ & $2.05$ & $3.07$ & $5.19$ & $21.73$ & $1.86$ & $\mathbf{0.55}$ & $23.70$ \\

  & {\ModelName (Pretrained)} & \cellcolor{g5_c4}$18292$ & $1.87$ & $5.49$ & $12.46$ & $35.55$ & $579.67$ & $\mathbf{1.00}$ & $1.42$ & $1.90$ & $2.41$ & $6.40$ & $41.14$ & $313.56$ & $14.13$ \\

  & {\ModelName (Finetuned)} & \cellcolor{g5_c3}$\mathbf{17192}$ & $\mathbf{1.41}$ & $\mathbf{2.28}$ & $\mathbf{3.67}$ & $\mathbf{7.07}$ & $\mathbf{42.03}$ & $\mathbf{1.00}$ & $\mathbf{1.00}$ & $\mathbf{1.25}$ & $\mathbf{1.46}$ & $\mathbf{2.33}$ & $41.14$ & $10.20$ & $14.13$ \\

  & {Optimal} & \cellcolor{g5_c2}$17160$ & $1.00$ & $1.00$ & $1.00$ & $1.00$ & $1.00$ & $1.00$ & $1.00$ & $1.00$ & $1.00$ & $1.00$ & $-$ & $-$ & $-$ \\
\hline

\multirow{6}{*}{ErgastF1}
  & {PG} & \cellcolor{g5_c1}$48185$ & $1.60$ & $4.26$ & $11.30$ & $27.47$ & $114.24$ & $\mathbf{1.00}$ & $1.32$ & $1.64$ & $1.98$ & $6.39$ & $-$ & $-$ & $-$ \\

  & {ALECE} & $50902$ & $1.75$ & $3.15$ & $5.18$ & $8.43$ & $53.81$ & $1.03$ & $1.32$ & $1.61$ & $2.08$ & $3.35$ & $233.34$ & $121.49$ & $7.06$ \\

  & {MSCN} & $45511$ & $10.20$ & $99.89$ & $353.02$ & $999.52$ & $1.32$$\cdot$$10^4$ & $1.52$ & $2.54$ & $4.92$ & $8.61$ & $13.72$ & $\mathbf{1.66}$ & $\mathbf{5.08}$ & $\mathbf{0.81}$ \\


  & {\ModelName (Pretrained)} & \cellcolor{g5_c4}$44532$ & $\mathbf{1.43}$ & $3.12$ & $6.45$ & $14.57$ & $67.97$ & $\mathbf{1.00}$ & $1.15$ & $1.46$ & $1.73$ & $\mathbf{2.24}$ & $41.14$ & $313.56$ & $7.14$ \\

  & {\ModelName (Finetuned)} & \cellcolor{g5_c3}$\mathbf{44350}$ & $\mathbf{1.43}$ & $\mathbf{2.46}$ & $\mathbf{4.06}$ & $\mathbf{7.37}$ & $\mathbf{29.96}$ & $\mathbf{1.00}$ & $\mathbf{1.00}$ & $\mathbf{1.42}$ & $\mathbf{1.57}$ & $2.28$ & $41.14$ & $5.21$ & $7.14$ \\

  & {Optimal} &\cellcolor{g5_c2}$44256$ & $1.00$ & $1.00$ & $1.00$ & $1.00$ & $1.00$ & $1.00$ & $1.00$ & $1.00$ & $1.00$ & $1.00$ & $-$ & $-$ & $-$ \\
\hline

\multirow{7}{*}{\shortstack{Visual \\ Genome}}
  & {PG} & \cellcolor{g5_c1}$6165$ & $1.17$ & $4.34$ & $15.12$ & $383.09$ & $1.17$$\cdot$$10^4$ & $1.04$ & $1.04$ & $1.44$ & $2.26$ & $4.72$ & $-$ & $-$ & $-$ \\

  & {ALECE} & $5874$ & $1.13$ & $1.34$ & $1.44$ & $1.58$ & $2.01$ & $\mathbf{1.00}$ & $\mathbf{1.00}$ & $1.04$ & $1.04$ & $\mathbf{1.08}$ & $232.72$ & $49.68$ & $6.25$ \\

  & {MSCN} & $12070$ & $2.84$ & $9.34$ & $18.91$ & $47.45$ & $108.48$ & $2.68$ & $2.74$ & $4.38$ & $4.39$ & $4.40$ & $\mathbf{1.54}$ & $39.86$ & $\mathbf{0.78}$ \\

  & {NeuroCard} & $6673$ & $1.04$ & $12.23$ & $14.35$ & $26.75$ & $53.75$ & $1.00$ & $1.20$ & $1.21$ & $1.28$ & $2.73$ & $38.38$ & $12.91$ & $9.85$ \\

  & {\ModelName (Pretrained)} & \cellcolor{g5_c4}$5840$ & $1.65$ & $3.59$ & $5.17$ & $15.67$ & $115.61$ & $\mathbf{1.00}$ & $\mathbf{1.00}$ & $1.01$ & $1.42$ & $2.64$ & $41.14$ & $313.56$ & $6.22$ \\

  & {\ModelName (Finetuned)} & \cellcolor{g5_c3}$\mathbf{5838}$ & $\mathbf{1.06}$ & $\mathbf{1.19}$ & $\mathbf{1.29}$ & $\mathbf{1.35}$ & $\mathbf{1.54}$ & $\mathbf{1.00}$ & $\mathbf{1.00}$ & $\mathbf{1.00}$ & $\mathbf{1.00}$ & $\mathbf{1.08}$ & $41.14$ & $\mathbf{11.48}$ & $6.22$ \\

  & {Optimal} & \cellcolor{g5_c2}$5834$ & $1.00$ & $1.00$ & $1.00$ & $1.00$ & $1.00$ & $1.00$ & $1.00$ & $1.00$ & $1.00$ & $1.00$ & $-$ & $-$ & $-$ \\
\hline

\end{tabular}

} 
\end{table*}
\end{scriptsize}

\stitle{Datasets.}
We selected four real-world datasets---IMDB, STATS, ErgastF1 and VisualGenome---for evaluation. They are widely adopted to examine the performance of CardEst methods in the literature~\cite{DBLP:journals/pvldb/NegiWKTMMKA23, DBLP:journals/pvldb/HanWWZYTZCQPQZL21, hilprecht_zero-shot_2022}. The detailed properties of these datasets are presented in Table~\ref{tab:datasets}. STATS~\cite{STATS_url,DBLP:journals/pvldb/HanWWZYTZCQPQZL21} is an anonymized compilation of all user-contributed content from the Stats Stack Exchange network~\cite{stats_web_url}. The associated query workload STATS-CEB contains 146 queries for testing. The IMDB dataset~\cite{leis2015good} is on movies and stars with the simplified JOB-light workload~\cite{job_light_url} containing 70 realistic queries. ErgastF1~\cite{ergast_url} provides extensive information on Formula 1 races spanning from the 1950 season to the present day. VisualGenome~\cite{genome_url} encompasses dense annotations of objects, attributes, and relationships within various genes. We use the method in Section~\ref{sec:training} to generate 148 and 186 queries on ErgastF1 and VisualGenome for testing, respectively.

\stitle{Competing Methods.}
We select the competing CardEst methods based on their performance reported in various benchmarks~\cite{DBLP:journals/pvldb/HanWWZYTZCQPQZL21,DBLP:journals/pvldb/SunZSLT21} and evaluation studies~\cite{DBLP:journals/pvldb/ZhuWHZPQZC21,DBLP:journals/corr/abs-2012-14743,DBLP:journals/pvldb/YangKLLDCS20}. For each category, namely traditional, data-driven, query-driven and hybrid CardEst methods, we select the representative ones as follows:

1) \textbf{PG} stands for the basic 1-D histogram based approach used in PostgreSQL~\cite{postgresql} for CardEst. This simple method serves as a baseline to measure the effectiveness of more sophisticated methods.

2) \textbf{NeuroCard}~\cite{DBLP:journals/pvldb/YangKLLDCS20} builds a single deep auto-regression model on the joint PDF of all tables in the database. The cardinality of any query could then be accessed from this model.

3) \textbf{DeepDB}~\cite{DBLP:journals/pvldb/HilprechtSKMKB20} learns the joint PDF of data using Sum-Product Networks (SPNs)~\cite{DBLP:conf/uai/PoonD11}. It builds several SPNs, each covering a subset of tables for CardEst.

4) \textbf{FactorJoin}~\cite{DBLP:journals/pacmmod/WuNAKM23} applies a cohesive factor graph to combine the learned cardinality on every single table, and the histograms of joins together for CardEst on multiple tables.

5) \textbf{MSCN}~\cite{DBLP:conf/cidr/KipfKRLBK19} utilizes a multi-set convolutional network to amalgamate data and query information together to learn the cardinality.

6) \textbf{ALECE}~\cite{li_alece_2023} uses a novel approach by meticulously featurizing data and SQL queries and learning to discover the implicit relationships between queries and underlying data using attention mechanisms.

For these methods, we apply their open-source implementations~\cite{neurocard_codes_url, deepdb_codes_url, FactorJoin_codes_url, mscn_codes_url, alece_codes_url} and follow their original configurations for parameter setting. For DeepDB, we set its RDC independence threshold to 0.3 and split each SPN node with at least 1\% of the input data; for NeuroCard, the sampling size is set to 8,000. We do not execute DeepDB and FactorJoin on the ErgastF1 and VisualGenome datasets as DeepDB only supports PK-FK joins, and the implementation of Factorjoin is hard-coded for the IMDB and STATS datasets.

Besides them, an \textbf{Optimal} approach, which utilizes the true cardinality for query plan generation, is also included for benchmarking. As shown in~\cite{leis2015good}, this method could find the optimal plans for most of the queries. Therefore, it could serve as a borderline to exhibit the best performance that the CardEst could bring for query optimization. We inject all CardEst methods into PostgreSQL~13.1 using PilotScope~\cite{10.14778/3641204.3641209}---a middleware for deploying ML algorithms into databases---for end-to-end testing in actual DBMS.

\begin{figure*} 
\includegraphics[width=\linewidth]{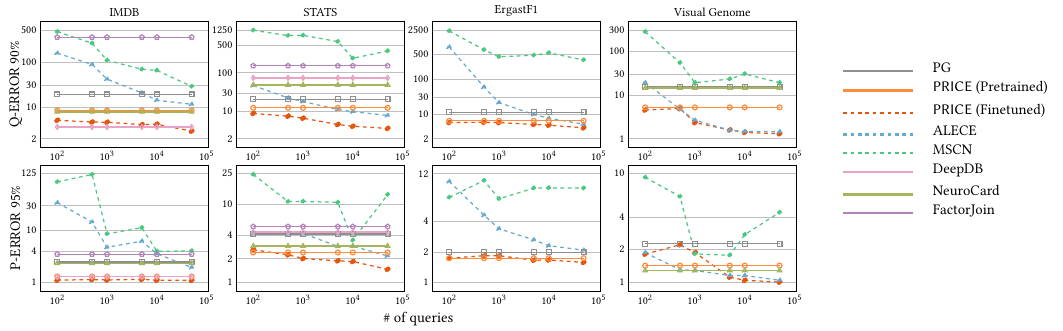}
\vspace{-20pt}
\captionsetup{font=large}
\caption{Estimation errors of cardinality estimation models across varying numbers of training queries.}
\label{fig:finetune}
\end{figure*} 

\stitle{Evaluation Metrics.}
We evaluate the effectiveness of each CardEst method by the following metrics: 

1) \textbf{End-to-end (E2E) Time} refers to the total time cost of plan generation using estimated cardinalities and physical plan execution. It reflects the performance gain that a CardEst method could bring to the query optimization process. Therefore, it is regarded as the gold standard metric to gauge the effectiveness of CardEst methods~\cite{DBLP:journals/pvldb/HanWWZYTZCQPQZL21}.

2) \textbf{Q-ERROR}~\cite{DBLP:journals/pvldb/MoerkotteNS09} assesses the relative multiplicative deviation of the estimation results from the exact value as $\text{Q-ERROR} = \max(
\widehat{\Card}(Q)/\Card(Q), \Card(Q)/\widehat{\Card}(Q))$.
It could reflect the accuracy of estimation results in a coarse-grained manner. However, it can not distinguish the cases of under-estimate and over-estimate and the impacts of estimation errors of different sub-queries w.r.t.~the plan generation.

3) \textbf{P-ERROR}~\cite{DBLP:journals/pvldb/HanWWZYTZCQPQZL21} corrects the drawbacks of Q-ERROR and could be computed without the actual execution of queries. By~\cite{leis2015good}, the plan cost estimated by the cost model is a good surrogate for the execution time. Therefore, we evaluate the quality of a CardEst method by comparing the execution cost of the plans derived from the estimated and true cardinalities. Specifically, for a query $Q$, let $\mathrm{C}^E$ and $\mathrm{C}^T$ denote the estimated and true cardinalities of all sub-queries of $Q$. Let $\mathrm{P}^E$ and $\mathrm{P}^T$ be the plan generated by the query optimizer by feeding $\mathrm{C}^E$ and $\mathrm{C}^T$ into it. 
Let $M$ be the cost model of the query optimizer and $M(P, C)$ denote the estimated cost of plan $P$ with cardinality $C$. $M(P, C)$ often approximates the execution time of plan $P$. 
Obviously, $M(\mathrm{P}^T, \mathrm{C}^T)$ is the estimated cost of executing the (most possible) optimal plan $\mathrm{P}^T$. In the actual execution of plan $\mathrm{P}^E$, the exact cardinality $\mathrm{C}^T$ would be instantiated, so the estimated execution cost of plan $\mathrm{P}^E$ is $M(\mathrm{P}^E, \mathrm{C}^T)$. Thus, we define it as $\text{P-ERROR} = M(\mathrm{P}^E, \mathrm{C}^T) / M(\mathrm{P}^T, \mathrm{C}^T)$. As analyzed in~\cite{DBLP:journals/pvldb/HanWWZYTZCQPQZL21}, P-ERROR considers the impacts of estimation errors of all sub-queries in terms of plan optimization and is more highly correlated to the E2E time than Q-ERROR.

Besides them, we also consider the \textbf{Training Time} for model training, the \textbf{Inference Time} for accessing cardinality from the CardEst models, and the \textbf{Model Size} for the amount of memory to store model parameters. These metrics could reflect the performance of each CardEst method from multiple views.

\vspace{-10pt}
\subsection{Performance on Unseen Datasets}
\label{sec:exp-pretrain}

We compare our \ModelName and other CardEst methods on the four unseen new databases. For fairness, we also apply ~$5\times 10^4$ queries on each dataset to train MSCN and ALECE. As shown in Table~\ref{fig:pretrain}, we have the following observations:

1) Our pretrained model \ModelName consistently outperforms other CardEst methods in terms of effectiveness. Specifically, for the most crucial E2E time, \ModelName surpasses the basic PG and other ML-based methods in almost all cases (except IMDB, where all methods attain good results). 
On STATS, the E2E time of \ModelName is $1.67\times$, $1.25\times$, and $1.41\times$ faster than PG, DeepDB, and ALECE, respectively. In terms of the Q-ERROR and P-ERROR metrics, \ModelName is always better than PG and has no significant gaps with other ML-based methods. This verifies the success of the design choices of our \ModelName model and its pretraining strategies.

2) Our \ModelName consistently demonstrates strong and stable performance across all datasets, while existing ML-based methods may exhibit significant variability. For instance, FactorJoin achieves near-optimal performance on IMDB and performs even worse than PG on STATS. This property of \ModelName plays an important role when actually deployed into DBMS.  By~\cite{10.14778/3641204.3641205}, the DBMS not only focuses on performance gains but also pays attention to avoiding performance regressions. The reason for that is our model encounters a number of different data distributions and query workloads during pretraining, so the learned parameters could generalize to attain stable performance on any database. Whereas, existing ML-based methods often have prior assumptions on the data distributions (e.g., DeepDB assumes attributes are not highly correlated), so they may fail when the dataset does not obey such assumptions. 

3) \ModelName is time and space efficient. Its inference time spans from several to a dozen milliseconds, which would not degrade the plan generation process. In the actual deployment, we only need to store one singleton \ModelName model having $41.14$MB for all databases. This is much smaller than storing the models trained by ALECE, NeuroCard, and DeepDB on each specific database. The pretraining cost of \ModelName is longer than other ML-based methods. However, we just need to train it once, and then the model can be seamlessly applied to any database.

In short, this set of experiments verifies the superiority of our \ModelName over other CardEst methods. It generates plans with much higher and more stable quality, incurs less time and space overheads, and exhibits the easiness for system deployment.

\vspace{-10pt}
\subsection{Performance of \ModelName with Finetuning}
\label{sec:exp-finetune}

We further examine the performance of our \ModelName after finetuning. Figure~\ref{fig:finetune} illustrates the estimation errors (the $90\%$-quantile Q-ERROR and $95\%$-quantile P-ERROR) of different CardEst models by varying the number of training queries. We omit the results of other metrics, such as E2E time, since they exhibit similar trends. PG, data-driven methods (DeepDB, NeuroCard and FactorJoin) and our pretrained model exhibit a stable horizontal line since their performance is independent of the training workload. For the finetuned version of \ModelName, we apply the corresponding training queries to tune the model further to fit each specific dataset. The results reported in Table~\ref{fig:pretrain} are obtained by the models finetuned (or trained) with $5 \times 10^4$ queries. Based on Table~\ref{fig:pretrain} and Figure~\ref{fig:finetune}, several key observations emerge:

1) \ModelName can be further enhanced to achieve near-optimal performance by finetuning. In Table~\ref{fig:pretrain}, our finetuned \ModelName exhibits the smallest Q-ERROR and P-ERROR values across most quantiles.
On most datasets, the P-ERROR values even approach $1.0$, the lower bound of this metric. In terms of the E2E time, our finetuned \ModelName achieves the shortest time in all four test datasets. The relative deviation to the optimal cases is all less than $0.4\%$.

2) The finetuning cost of \ModelName is small and acceptable. From Figure~\ref{fig:finetune}, our \ModelName could outperform almost all other methods in terms of Q-ERROR and P-ERROR when finetuning with even 100 queries. Collecting and executing such a few training queries takes less than one hour. The finetuning time to tune the model of \ModelName is only around 10 minutes. Even if we collect and execute $5 \times 10^4$ queries by around 216 hours (or 9 days) and consume several hours to train models of MSCN and ALECE, their performance is still worse than \ModelName. This is because the existing query-driven and hybrid CardEst methods need to be learned from scratch for each specific database. Whereas, for \ModelName, the model has already captured the general paradigm for mapping joint PDFs and query information to the cardinality after pretraining, 
so we could tune the model to fit a specific database with a small number of queries.

\begin{figure}

\begin{tikzpicture}

\begin{groupplot}[
    group style={
        group size=2 by 1, 
        vertical sep=0.85cm, 
        horizontal sep=1.0cm, 
        group name= fig_1248x, 
    },
    width=4.8cm, 
    height=3.6cm, 
    xticklabel style={font=\scriptsize},
    yticklabel style={font=\scriptsize},
    xlabel style={font=\footnotesize, yshift=8pt},
    ylabel style={font=\footnotesize, yshift=-18pt},
    ymajorgrids=true,
    grid style=solid,
    xtick style={draw=none},
    ytick style={draw=none},
    xtick={1, 2, 3, 4},
    xticklabels={1x, 2x, 4x, 8x},
    enlarge x limits={abs=0.35cm},
    enlarge y limits={abs=0.15cm},
]

\nextgroupplot[
    ylabel={Q-ERROR $90\%$},
    ymode=log,
    ymin=5, ymax=1250,
    ytick={5, 20, 100, 400, 1250},
    yticklabels={5, 20, 100, 400, 1250},
]
\addplot[
    color=g3_c1, 
    mark=square,
    mark size=1.5pt, 
    line width=1pt, 
    error bars/.cd,
    y dir=both,
    y explicit,
] coordinates {
    (1, 18.12)
    (2, 14.90)
    (3, 16.69)
    (4, 20.71)
};

\addplot[
    color=g3_c3, 
    dashed,
    mark=triangle*,
    mark size=1.5pt, 
    line width=1pt, 
    error bars/.cd,
    y dir=both,
    y explicit,
] coordinates {
    (1, 6.67)
    (2, 9.95)
    (3, 20.52)
    (4, 69.31)
};

\addplot[
    color=g3_c4, 
    dashed,
    mark=diamond*,
    mark size=1.5pt, 
    line width=1pt, 
    error bars/.cd,
    y dir=both,
    y explicit,
] coordinates {
    (1, 87.52)
    (2, 113.52)
    (3, 300.81)
    (4, 962.73)
};

\addplot[
    color=g3_c5, 
    mark=diamond*,
    mark size=1.5pt, 
    line width=1pt, 
    error bars/.cd,
    y dir=both,
    y explicit,
] coordinates {
    (1, 40.81)
    (2, 128.68)
    (3, 406.56)
    (4, 1283.90)
};

\addplot[
    color=g3_c6, 
    mark=triangle,
    mark size=1.5pt, 
    line width=1pt, 
    error bars/.cd,
    y dir=both,
    y explicit,
] coordinates {
    (1, 42.52)
    (2, 115.72)
    (3, 333.23)
    (4, 1230.80)
};

\addplot[
    color=g3_c7, 
    mark=pentagon,
    mark size=1.5pt, 
    line width=1pt, 
    error bars/.cd,
    y dir=both,
    y explicit,
] coordinates {
    (1, 120.36)
    (2, 57.05)
    (3, 44.61)
    (4, 70.81)
};

\addplot[
    color=g3_c2, 
    mark=o,
    mark size=1.5pt, 
    line width=1pt, 
    error bars/.cd,
    y dir=both,
    y explicit,
] coordinates {
    (1, 12.37)
    (2, 10.81)
    (3, 14.37)
    (4, 12.46)
};

\addplot[
    color=g3_c8, 
    dashed,
    mark=*,
    mark size=1.5pt, 
    line width=1pt, 
    error bars/.cd,
    y dir=both,
    y explicit,
] coordinates {
    (1, 6.29)
    (2, 5.39)
    (3, 7.08)
    (4, 6.13)
};

\nextgroupplot[
    ylabel style={yshift=-2pt},
    ylabel={P-ERROR $95\%$},
    ymode=log,
    ymin=2, ymax=30,
    ytick={2, 4, 8, 16, 30},
    yticklabels={2, 4, 8, 16, 30}
]
\addplot[
    color=g3_c1, 
    mark=square,
    mark size=1.5pt, 
    line width=1pt, 
    error bars/.cd,
    y dir=both,
    y explicit,
] coordinates {
    (1, 3.02)
    (2, 3.20)
    (3, 2.57)
    (4, 4.16)
};

\addplot[
    color=g3_c3, 
    dashed,
    mark=triangle*,
    mark size=1.5pt, 
    line width=1pt, 
    error bars/.cd,
    y dir=both,
    y explicit,
] coordinates {
    (1, 3.13)
    (2, 2.80)
    (3, 4.31)
    (4, 4.55)
};

\addplot[
    color=g3_c4, 
    dashed,
    mark=diamond*,
    mark size=1.5pt, 
    line width=1pt, 
    error bars/.cd,
    y dir=both,
    y explicit,
] coordinates {
    (1, 6.20)
    (2, 8.11)
    (3, 8.38)
    (4, 9.38)
};

\addplot[
    color=g3_c5, 
    mark=diamond*,
    mark size=1.5pt, 
    line width=1pt, 
    error bars/.cd,
    y dir=both,
    y explicit,
] coordinates {
    (1, 6.35)
    (2, 9.14)
    (3, 22.91)
    (4, 28.46)
};

\addplot[
    color=g3_c6, 
    mark=triangle,
    mark size=1.5pt, 
    line width=1pt, 
    error bars/.cd,
    y dir=both,
    y explicit,
] coordinates {
    (1, 8.18)
    (2, 7.22)
    (3, 5.29)
    (4, 6.42)
};

\addplot[
    color=g3_c7, 
    mark=pentagon,
    mark size=1.5pt, 
    line width=1pt, 
    error bars/.cd,
    y dir=both,
    y explicit,
] coordinates {
    (1, 5.70)
    (2, 7.99)
    (3, 12.48)
    (4, 19.82)
};

\addplot[
    color=g3_c2, 
    mark=o,
    mark size=1.5pt, 
    line width=1pt, 
    error bars/.cd,
    y dir=both,
    y explicit,
] coordinates {
    (1, 4.51)
    (2, 4.01)
    (3, 2.13)
    (4, 2.41)
};

\addplot[
    color=g3_c8, 
    dashed,
    mark=*,
    mark size=1.5pt, 
    line width=1pt, 
    error bars/.cd,
    y dir=both,
    y explicit,
] coordinates {
    (1, 2.39)
    (2, 2.38)
    (3, 1.98)
    (4, 2.36)
};

\end{groupplot}

\matrix[
  matrix of nodes,
  anchor=north,
  every node/.style={font=\scriptsize, anchor=west} 
] at ($(fig_1248x c1r1.south)!0.47!(fig_1248x c2r1.south)$)[below, yshift=-1em]
{
  \tikz \draw[g3_c1, thick, line width=2pt] (0, 0) -- (0.5, 0); \ PG &
  \tikz \draw[g3_c2, thick, line width=2pt] (0, 0) -- (0.5, 0); \ \ModelName (Pretrained) &
  \tikz \draw[g3_c8, thick, densely dashed, line width=2pt] (0, 0) -- (0.5, 0); \ \ModelName (Finetuned) &
  \tikz \draw[g3_c3, thick, densely dashed, line width=2pt] (0, 0) -- (0.5, 0); \  ALECE \\
  \tikz \draw[g3_c4, thick, densely dashed, line width=2pt] (0, 0) -- (0.5, 0); \ MSCN &
  \tikz \draw[g3_c5, thick, line width=2pt] (0, 0) -- (0.5, 0); \ DeepDB &
  \tikz \draw[g3_c6, thick, line width=2pt] (0, 0) -- (0.5, 0); \ NeuroCard &
  \tikz \draw[g3_c7, thick, line width=2pt] (0, 0) -- (0.5, 0); \  FactorJoin \\
};
\end{tikzpicture}
\vspace{-20pt} 
\captionsetup{font=large}
\caption{Estimation errors of cardinality estimation models during data updating processes.}
\label{fig:gen_updates}
\vspace{-5pt} 
\end{figure}
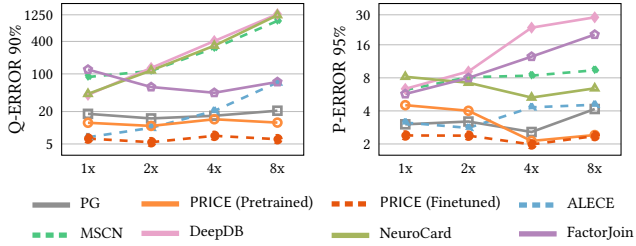
\begin{figure}

\begin{tikzpicture}

\begin{groupplot}[
    group style={
        group size=2 by 1, 
        vertical sep=0.85cm, 
        horizontal sep=1.0cm, 
        group name= fig_1248x, 
    },
    width=4.8cm, 
    height=3.6cm, 
    xticklabel style={font=\scriptsize},
    yticklabel style={font=\scriptsize},
    xlabel style={font=\footnotesize, yshift=8pt},
    ylabel style={font=\footnotesize, yshift=-18pt},
    xmode=log,
    xtick={100, 1000, 10000, 100000},
    xticklabels={$10^2$, $10^3$, $10^4$, $10^5$},
    ymajorgrids=true,
    grid style=solid,
    xtick style={draw=none},
    ytick style={draw=none},
    enlarge x limits={abs=0.35cm},
    enlarge y limits={abs=0.15cm},
]

\nextgroupplot[
    xlabel={\# of queries},
    ylabel={Q-ERROR $50\%$},
    ymin=1.25, ymax=2,
    ytick={1.25, 1.5, 1.75, 2},
    yticklabels={1.25, 1.50, 1.75, 2.00},
]
\addplot[
    color=g6_c1, 
    mark=*,
    mark size=1.5pt, 
    line width=1pt, 
    error bars/.cd,
    y dir=both,
    y explicit,
] coordinates {
    (100, 1.74)
    (500, 1.72)
    (1000, 1.81)
    (5000, 1.58)
    (10000, 1.52)
    (50000, 1.43)
};

\addplot[
    color=g6_c2, 
    mark=*,
    mark size=1.5pt, 
    line width=1pt, 
    error bars/.cd,
    y dir=both,
    y explicit,
] coordinates {
    (100, 1.85)
    (500, 1.77)
    (1000, 1.75)
    (5000, 1.55)
    (10000, 1.45)
    (50000, 1.41)
};

\addplot[
    color=g6_c3, 
    mark=square*,
    mark size=1.5pt, 
    line width=1pt, 
    error bars/.cd,
    y dir=both,
    y explicit,
] coordinates {
    (100, 1.87)
    (500, 1.87)
    (1000, 1.87)
    (5000, 1.87)
    (10000, 1.87)
    (50000, 1.87)
};

\addplot[
    color=g6_c4, 
    mark=*,
    mark size=1.5pt, 
    line width=1pt, 
    error bars/.cd,
    y dir=both,
    y explicit,
] coordinates {
    (100, 1.87)
    (500, 1.87)
    (1000, 1.87)
    (5000, 1.87)
    (10000, 1.87)
    (50000, 1.87)
};

\nextgroupplot[
    xlabel={\# of queries},
    ylabel style={yshift=-2pt},
    ylabel={Q-ERROR $90\%$},
    ymin=2, ymax=22,
    ytick={2, 6, 10, 14, 18, 22},
    yticklabels={2, 6, 10, 14, 18, 22}
]
\addplot[
    color=g6_c1, 
    mark=*,
    mark size=1.5pt, 
    line width=1pt, 
    error bars/.cd,
    y dir=both,
    y explicit,
] coordinates {
    (100, 8.17)
    (500, 8.56)
    (1000, 7.59)
    (5000, 7.01)
    (10000, 5.20)
    (50000, 5.09)
};

\addplot[
    color=g6_c2, 
    mark=*,
    mark size=1.5pt, 
    line width=1pt, 
    error bars/.cd,
    y dir=both,
    y explicit,
] coordinates {
    (100, 8.92)
    (500, 7.58)
    (1000, 6.72)
    (5000, 4.63)
    (10000, 4.16)
    (50000, 3.67)
};

\addplot[
    color=g6_c3, 
    mark=square*,
    mark size=1.5pt, 
    line width=1pt, 
    error bars/.cd,
    y dir=both,
    y explicit,
] coordinates {
    (100, 20.71)
    (500, 20.71)
    (1000, 20.71)
    (5000, 20.71)
    (10000, 20.71)
    (50000, 20.71)
};

\addplot[
    color=g6_c4, 
    mark=*,
    mark size=1.5pt, 
    line width=1pt, 
    error bars/.cd,
    y dir=both,
    y explicit,
] coordinates {
    (100, 12.46)
    (500, 12.46)
    (1000, 12.46)
    (5000, 12.46)
    (10000, 12.46)
    (50000, 12.46)
};

\end{groupplot}

\matrix[
  matrix of nodes,
  anchor=north,
  every node/.style={font=\footnotesize, anchor=west} 
] at ($(fig_1248x c1r1.south)!0.47!(fig_1248x c2r1.south)$)[below, yshift=-1.5em]
{
  \tikz \draw[g6_c3, thick, line width=2pt] (0, 0) -- (1, 0); \ PG &
  \tikz \draw[g6_c4, thick, line width=2pt] (0, 0) -- (1, 0); \ Pretrained \\
  
  \tikz \draw[g6_c1, thick, line width=2pt] (0, 0) -- (1, 0); \ Sub-Database Finetuned &
  \tikz \draw[g6_c2, thick, line width=2pt] (0, 0) -- (1, 0); \ Full-Database Finetuned \\
};
\end{tikzpicture}
\vspace{-20pt} 
\captionsetup{font=large}
\caption{\ModelName finetuned on the Sub-Database and the Full-Database workload and evaluated on the Full-Database.}
\label{fig:gen_scaling}
\vspace{-15pt} 
\end{figure}
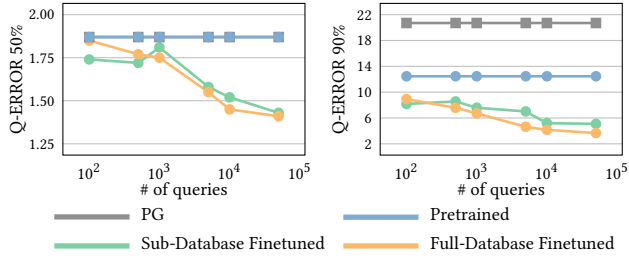

\subsection{Generalization Ability of \ModelName}
\label{sec:exp-robust}

In this section, we examine the generalization ability of our pretrained model \ModelName. We conduct experiments in three scenarios commonly occurring in real-world applications, namely data updates, data scaling and query workload drifts. 

\stitle{Generalization Ability to Data Updates.}
In real-world DBMS, data is constantly updated. We apply the STATS dataset with timestamps to test the performance of \ModelName on updated data. Similar to~\cite{li_alece_2023,DBLP:journals/pvldb/HanWWZYTZCQPQZL21}, we split the STATS datasets according to the creation time of tuples and obtain 4 datasets containing $1/8$, $2/8$, $4/8$ and $8/8$ tuples of the full data, respectively. We denote them as STATS $1$x, $2$x, $4$x to $8$x. For our \ModelName, we directly test the pretrained model on the 4 datasets. For each existing CardEst method and the finetuned version of \ModelName, we apply the underlying data and execute the query workload on STATS $1$x (12.5\% of the full data) for collecting workload statistics and model training (finetuning for \ModelName). Then, we use the same test queries to evaluate its performance on STATS $1$x, $2$x, $4$x to $8$x. Notably, although the test queries remain the same on the 4 datasets, the true cardinalities differ due to data changes. We only report the results of $90\%$-quantile Q-ERROR and $95\%$-quantile P-ERROR and omit the similar results of other metrics.

From Figure~\ref{fig:gen_updates}, we find existing CardEst methods (except the basic PG) exhibit diminished performance when faced with data updates, and DeepDB, NeuroCard, and MSCN show notable performance declines. It also resembles their behaviors in Section~\ref{sec:exp-pretrain} on unseen datasets. These models are specifically trained to fit each dataset, which may fall down on updated data with different distributions.

However, both the original pretrained and finetuned \ModelName maintain a promising and stable performance across all datasets. These results align with our observations in Section~\ref{sec:exp-pretrain}. The pretrained \ModelName is trained to capture the meta-knowledge for CardEst, so it easily adapts to arbitrary datasets, including both unseen and updated data. For the finetuned \ModelName, although we finetune it on STATS $1$x, it could also adjust its parameters to capture some coarse-grained knowledge on the data distributions of the STATS dataset. As a result, when it is transferred to STATS $2$x, $4$x to $8$x, the performance can still be further improved.

\stitle{Generalization Ability to Data Scaling.}
Later, we present more details on the impact of different data volumes on the finetuned \ModelName. We finetune \ModelName on STATS $1$x (partial of the dataset, denoted as Sub-Database) with a different number of queries and compare its performance with the model finetuned on the Full-Database (STATS $8$x). Figure~\ref{fig:gen_scaling} exhibits a surprising result: the performance of \ModelName finetuned on the partial and full dataset exhibits comparable performance. No matter how many training queries are applied, their performance is very close and consistently better than basic PG and original \ModelName without finetuning. This further verifies the generalization ability of \ModelName, which just focuses on the data distribution but not the data volume. This property is very appealing as executing the query workload on a smaller dataset is much faster, so we could consume much less time to collect statistics on a new dataset for finetuning. In the above setting, in comparison to executing the query workload on the full dataset for finetuning, we consume only $1.5\%$ of time on STATS $1$x.

\begin{figure}
\vspace{-5pt}
\includegraphics[width=\linewidth]{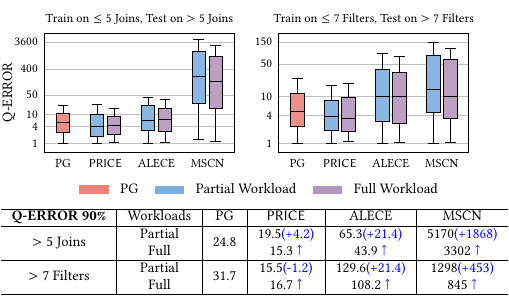}
\vspace{-20pt} 
\captionsetup{font=large}
\caption{Estimation errors of cardinality estimation models across workload drifts.}
\label{fig:gen_workload}
\vspace{-15pt} 
\end{figure}

\stitle{Generalization Ability to Query Workload Drifts.}
In actual \break DBMS, the query workloads may change from time to time~\cite{10.14778/3641204.3641205}. As data-driven CardEst methods are not sensitive to workload drifts~\cite{DBLP:journals/pvldb/HanWWZYTZCQPQZL21,DBLP:journals/pvldb/ZhuWHZPQZC21}, we only compare our \ModelName with ALECE and MSCN. For the training workload on a dataset, we obtain two partial workloads containing only no more than $5$ joins and $7$ filtering predicates, respectively. \ModelName is pretrained on each partial workload derived from the 26 datasets (mentioned in Section~\ref{sec:training}) and tested on STATS. ALECE and MSCN are directly trained on each partial workload of STATS. For testing, we only employ queries from the test workload containing more than $5$ joins or $7$ filtering predicates to examine their performance on different queries. Figure~\ref{fig:gen_workload} reports their performance on training (or pretraining for \ModelName) with partial and full workloads. We have the following observations:

1) \ModelName exhibits promising generalization ability to workload drifts. Despite training on the partial workload, \ModelName shows minimal performance degradation (the blue numbers in Figure~\ref{fig:gen_workload}) compared to training on the full workload. This validates the effectiveness of the attention-based architecture in handling queries with different numbers of joins and filtering predicates. Similarly, another attention-based approach, ALECE, also maintains relatively stable performance. Its Q-ERROR remains consistent, but is much higher than \ModelName, indicating the potential limitations in handling complex query patterns. For MSCN, which uses simple deep neural networks, the performance drops very significantly.

2) For complex queries with a relatively large number of joins or filtering predicates, only our \ModelName could obtain better performance than the basic PG. On the contrary, both ALECE and MSCN perform very poorly on such complex queries (consistently worse than PG), no matter whether they are trained on full or partial workloads. This is due to the different learning mechanisms between \ModelName and existing methods. \ModelName aims to learn the general approach to combine and correct the simple distributions on every single table (encoded as features) to the joint PDF, making it easily applicable to any number of joins and filtering predicates. Whereas, the models in ALECE and MSCN learn the specific parameters to map queries to their cardinality. As the training workload may not contain a sufficient number of complex queries (executing them is time-consuming), the models may not be well trained to capture such mapping relations.

\begin{figure}

\begin{tikzpicture}

\begin{groupplot}[
    group style={
        group size=2 by 2, 
        vertical sep=0.85cm, 
        horizontal sep=1.0cm, 
        group name= fig_num, 
    },
    width=4.8cm, 
    height=3.6cm, 
    xticklabel style={font=\scriptsize},
    yticklabel style={font=\scriptsize},
    xlabel style={font=\footnotesize, yshift=8pt},
    ylabel style={font=\footnotesize, yshift=-20pt},
    ymajorgrids=true,
    grid style=solid,
    xtick style={draw=none},
    ytick style={draw=none},
    enlarge x limits={abs=0.35cm},
    enlarge y limits={abs=0.15cm},
]

\nextgroupplot[
    xlabel={\# of datasets},
    xtick={5, 10, 15, 20, 26},
    ylabel={Q-ERROR $50\%$},
    ylabel style={yshift=-2pt},
    ymin=1, ymax=6,
    ytick={1, 2, 3, 4, 5, 6},
]
\addplot[
    color=g2_c1, 
    mark=*,
    mark size=1.5pt, 
    line width=1pt, 
    error bars/.cd,
    y dir=both,
    y explicit,
] coordinates {
    (5, 5.54)
    (10, 2.73)
    (15, 2.03)
    (20, 1.73)
    (26, 1.78)
};

\addplot[
    color=g2_c2, 
    mark=triangle*,
    mark size=1.5pt, 
    line width=1pt, 
    error bars/.cd,
    y dir=both,
    y explicit,
] coordinates {
    (5, 2.56)
    (10, 2.13)
    (15, 2.21)
    (20, 1.90)
    (26, 1.87)
};

\addplot[
    color=g2_c3, 
    mark=square*,
    mark size=1.5pt, 
    line width=1pt, 
    error bars/.cd,
    y dir=both,
    y explicit,
] coordinates {
    (5, 1.96)
    (10, 1.66)
    (15, 1.69)
    (20, 1.62)
    (26, 1.43)
};

\addplot[
    color=g2_c4, 
    mark=diamond*,
    mark size=1.5pt, 
    line width=1pt, 
    error bars/.cd,
    y dir=both,
    y explicit,
] coordinates {
    (5, 1.48)
    (10, 1.60)
    (15, 1.86)
    (20, 2.01)
    (26, 1.65)
};

\nextgroupplot[
    xlabel={\# of datasets},
    xtick={5, 10, 15, 20, 26},
    ylabel={Q-ERROR $90\%$},
    ymin=1, ymax=75,
    ytick={1, 15, 30, 45, 60, 75},
]
\addplot[
    color=g2_c1, 
    mark=*,
    mark size=1.5pt, 
    line width=1pt, 
    error bars/.cd,
    y dir=both,
    y explicit,
] coordinates {
    (5, 41.56)
    (10, 27.57)
    (15, 18.46)
    (20, 7.31)
    (26, 8.40)
};

\addplot[
    color=g2_c2, 
    mark=triangle*,
    mark size=1.5pt, 
    line width=1pt, 
    error bars/.cd,
    y dir=both,
    y explicit,
] coordinates {
    (5, 43.15)
    (10, 23.06)
    (15, 19.60)
    (20, 15.94)
    (26, 12.46)
};

\addplot[
    color=g2_c3, 
    mark=square*,
    mark size=1.5pt, 
    line width=1pt, 
    error bars/.cd,
    y dir=both,
    y explicit,
] coordinates {
    (5, 28.78)
    (10, 10.08)
    (15, 9.95)
    (20, 8.19)
    (26, 6.45)
};

\addplot[
    color=g2_c4, 
    mark=diamond*,
    mark size=1.5pt, 
    line width=1pt, 
    error bars/.cd,
    y dir=both,
    y explicit,
] coordinates {
    (5, 72.97)
    (10, 8.79)
    (15, 8.89)
    (20, 7.55)
    (26, 5.17)
};

\nextgroupplot[
    xlabel={\# of queries},
    xmode=log, 
    log basis x={10}, 
    xtick={100, 1000, 10000, 100000},
    xticklabels={$10^2$, $10^3$, $10^4$, $10^5$},
    ylabel={Q-ERROR $50\%$},
    ylabel style={yshift=-2pt},
    ymin=1, ymax=4,
    ytick={1, 2, 3, 4},
]
\addplot[
    color=g2_c1, 
    mark=*,
    mark size=1.5pt, 
    line width=1pt, 
    error bars/.cd,
    y dir=both,
    y explicit,
] coordinates {
    (100, 3.40)
    (500, 2.13)
    (1000, 2.06)
    (5000, 1.91)
    (10000, 1.78)
    (30000, 1.69)
    (50000, 1.78)
};

\addplot[
    color=g2_c2, 
    mark=triangle*,
    mark size=1.5pt, 
    line width=1pt, 
    error bars/.cd,
    y dir=both,
    y explicit,
] coordinates {
    (100, 2.55)
    (500, 1.88)
    (1000, 1.85)
    (5000, 1.98)
    (10000, 2.08)
    (30000, 1.96)
    (50000, 1.87)
};

\addplot[
    color=g2_c3, 
    mark=square*,
    mark size=1.5pt, 
    line width=1pt, 
    error bars/.cd,
    y dir=both,
    y explicit,
] coordinates {
    (100, 1.95)
    (500, 1.64)
    (1000, 1.65)
    (5000, 1.60)
    (10000, 1.60)
    (30000, 1.47)
    (50000, 1.43)
};

\addplot[
    color=g2_c4, 
    mark=diamond*,
    mark size=1.5pt, 
    line width=1pt, 
    error bars/.cd,
    y dir=both,
    y explicit,
] coordinates {
    (100, 2.92)
    (500, 1.36)
    (1000, 1.34)
    (5000, 1.32)
    (10000, 1.51)
    (30000, 1.74)
    (50000, 1.65)
};

\nextgroupplot[
    xlabel={\# of queries},
    xmode=log, 
    log basis x={10}, 
    xtick={100, 1000, 10000, 100000},
    xticklabels={$10^2$, $10^3$, $10^4$, $10^5$},
    ylabel={Q-ERROR $90\%$},
    ymin=1, ymax=55,
    ytick={1, 15, 30, 45, 55},
]
\addplot[
    color=g2_c1, 
    mark=*,
    mark size=1.5pt, 
    line width=1pt, 
    error bars/.cd,
    y dir=both,
    y explicit,
] coordinates {
    (100, 52.56)
    (500, 20.44)
    (1000, 18.99)
    (5000, 12.43)
    (10000, 7.81)
    (30000, 6.50)
    (50000, 8.40)
};

\addplot[
    color=g2_c2, 
    mark=triangle*,
    mark size=1.5pt, 
    line width=1pt, 
    error bars/.cd,
    y dir=both,
    y explicit,
] coordinates {
    (100, 26.37)
    (500, 21.22)
    (1000, 20.74)
    (5000, 19.85)
    (10000, 20.80)
    (30000, 18.10)
    (50000, 12.46)
};

\addplot[
    color=g2_c3, 
    mark=square*,
    mark size=1.5pt, 
    line width=1pt, 
    error bars/.cd,
    y dir=both,
    y explicit,
] coordinates {
    (100, 14.41)
    (500, 11.11)
    (1000, 11.44)
    (5000, 10.23)
    (10000, 9.12)
    (30000, 7.28)
    (50000, 6.45)
};

\addplot[
    color=g2_c4, 
    mark=diamond*,
    mark size=1.5pt, 
    line width=1pt, 
    error bars/.cd,
    y dir=both,
    y explicit,
] coordinates {
    (100, 34.71)
    (500, 18.42)
    (1000, 17.07)
    (5000, 5.74)
    (10000, 5.14)
    (30000, 4.32)
    (50000, 5.17)
};

\end{groupplot}

\matrix[
  matrix of nodes,
  anchor=north,
  every node/.style={font=\footnotesize, anchor=west} 
] at ($(fig_num c1r2.south)!0.47!(fig_num c2r2.south)$)[below, yshift=-2em]
{
  \tikz \draw[g2_c1, thick, line width=2pt] (0, 0) -- (0.5, 0); \ IMDB &
  \tikz \draw[g2_c2, thick, line width=2pt] (0, 0) -- (0.5, 0); \ STATS &
  \tikz \draw[g2_c3, thick, line width=2pt] (0, 0) -- (0.5, 0); \  ErgastF1 &
  \tikz \draw[g2_c4, thick, line width=2pt] (0, 0) -- (0.5, 0); \ Visual Genome \\
};
\end{tikzpicture}
\vspace{-10pt} 
\captionsetup{font=large}
\caption{Evaluation of \ModelName's performance with varied quantities of datasets and queries.}
\label{fig:exp-datasets}
\vspace{-20pt} 
\end{figure}
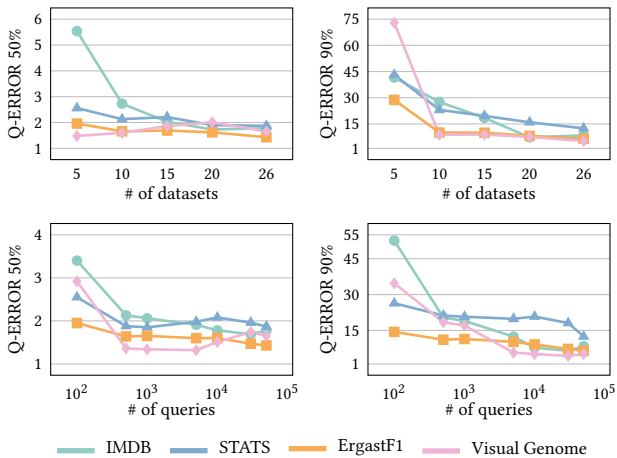
\vspace{-1pt}
\subsection{Impacts of Training Data to Pretrain \ModelName}
\label{sec:exp-dataset}

Collecting broader and richer training data from varied databases is crucial for developing a robust pretrained model. In this set of experiments, we explore how the quantity and diversity of datasets and the volume of training workload impact our pretrained \ModelName.


\stitle{Number of Training Datasets.}
Figure~\ref{fig:exp-datasets} shows the performance of \ModelName on unseen testing datasets that are pretrained using different numbers of datasets. We find that:

1) There is a clear trend that the $50\%$ and $90\%$-quantile Q-ERROR of \ModelName progressively improves when the number of datasets increases and eventually stabilizes after around $15$---$20$ datasets. The trends of other metrics are also similar, so we omit them due to space limits. It indicates that \ModelName can capture enough knowledge and transferable parameters for CardEst using not too many datasets. 

2) With a small number of datasets, the performance of \ModelName varies significantly in different datasets. Specifically, when trained using only 5 datasets, \ModelName performs well on ErgastF1 but poorly on IMDB. However, when trained over 15 datasets, its performance remains stable over all datasets (also witnessed in Section~\ref{sec:exp-pretrain}). This is because a small number of datasets are not diversified enough, so the knowledge learned by \ModelName may be biased and can not generalize well. This underscores the importance of selecting datasets from a variety of domains with diverse properties to achieve the high generalization capability of \ModelName.



\noindent\underline{\textbf{Number of Training Queries.}} Beyond the dataset quantity, the number of training queries collected from each dataset also directly affects the performance of the pretrained \ModelName. Therefore, we also pretrain \ModelName with a different numbers of queries on each dataset and evaluate its performance on four unseen test datasets.

From Figure~\ref{fig:exp-datasets}, we observe that the overall performance of \ModelName steadily improves when more queries are applied for training. It is natural as more training queries bring the \ModelName's model more knowledge on how to filter the joint PDF for computing cardinality. Additionally, the improvement of \ModelName
becomes marginal after $1.0 \times 10^3$ training queries, indicating the model has mastered enough knowledge. At that time, the workload contains only $2.6 \times 10^4$ queries to pretrain \ModelName. It is much cheaper than existing query-driven and hybrid CardEst methods, which require collecting a large volume of training queries on each dataset.
We note that, on the STATS dataset, ALECE is required to collect around the same volume of queries to attain comparable performance of \ModelName pretrained with $2.6 \times 10^4$ queries. For MSCN, even if we train it on $5 \times 10^4$ queries (results reported in Table~\ref{fig:pretrain}), its performance is still much worse than \ModelName pretrained with only $2.6 \times 10^4$ queries.

\vspace{-4pt}
\section{Conclusion and Future Work}
\label{sec:conclusion}

In this paper, we introduce \ModelName, a \underline{PR}etrained mult\underline{I}-table \underline{C}ard\underline{E}st model that overcomes the limitations of existing methods by enabling quick deployment to new, unseen databases with minimal preparatory steps. Inspired by pretrained NLP models, \ModelName uses low-level, transferable features to learn high-level representations for CardEst, achieving accurate estimation with minimal cost.
Pretraining on diverse datasets, \ModelName demonstrates strong generalization capabilities and superior performance compared to existing methods. After finetuning with minimal database-specific data, it achieves near-optimal performance. \ModelName also incurs lower time and space costs, making it a practical solution for DBMS.

For future enhancements of \ModelName, it is crucial to extend its capabilities to handle more complex query types (e.g., non-equal and non-inner joins, $\mathtt{LIKE}$ queries). 
Additionally, we plan to extend the pretraining paradigm to other DBMS tasks, including but not limited to cost estimation, index recommendation, and view advisor.


\clearpage

\bibliographystyle{ACM-Reference-Format}
\bibliography{ref}


\begin{thebibliography}{86}


\ifx \showCODEN    \undefined \def \showCODEN     #1{\unskip}     \fi
\ifx \showDOI      \undefined \def \showDOI       #1{#1}\fi
\ifx \showISBNx    \undefined \def \showISBNx     #1{\unskip}     \fi
\ifx \showISBNxiii \undefined \def \showISBNxiii  #1{\unskip}     \fi
\ifx \showISSN     \undefined \def \showISSN      #1{\unskip}     \fi
\ifx \showLCCN     \undefined \def \showLCCN      #1{\unskip}     \fi
\ifx \shownote     \undefined \def \shownote      #1{#1}          \fi
\ifx \showarticletitle \undefined \def \showarticletitle #1{#1}   \fi
\ifx \showURL      \undefined \def \showURL       {\relax}        \fi
\providecommand\bibfield[2]{#2}
\providecommand\bibinfo[2]{#2}
\providecommand\natexlab[1]{#1}
\providecommand\showeprint[2][]{arXiv:#2}

\bibitem[\protect\citeauthoryear{??}{ale}{tory}]%
        {alece_codes_url}
 \bibinfo{year}{ALECE Github repository}\natexlab{}.
\newblock \bibinfo{howpublished}{\url{https://github.com/pfl-cs/ALECE}}.
\newblock


\bibitem[\protect\citeauthoryear{??}{dee}{tory}]%
        {deepdb_codes_url}
 \bibinfo{year}{DeepDB Github repository}\natexlab{}.
\newblock \bibinfo{howpublished}{\url{https://github.com/DataManagementLab/deepdb-public}}.
\newblock


\bibitem[\protect\citeauthoryear{??}{erg}{aset}]%
        {ergast_url}
 \bibinfo{year}{ErgastF1 Dataset}\natexlab{}.
\newblock \bibinfo{howpublished}{\url{https://relational-data.org/dataset/ErgastF1}}.
\newblock


\bibitem[\protect\citeauthoryear{??}{Fac}{tory}]%
        {FactorJoin_codes_url}
 \bibinfo{year}{FactorJoin Github repository}\natexlab{}.
\newblock \bibinfo{howpublished}{\url{https://github.com/wuziniu/FactorJoin}}.
\newblock


\bibitem[\protect\citeauthoryear{??}{imd}{aset}]%
        {imdb_url}
 \bibinfo{year}{IMDB Dataset}\natexlab{}.
\newblock \bibinfo{howpublished}{\url{http://homepages.cwi.nl/~boncz/job/imdb.tgz}}.
\newblock


\bibitem[\protect\citeauthoryear{??}{job}{load}]%
        {job_light_url}
 \bibinfo{year}{Job-light Workload}\natexlab{}.
\newblock \bibinfo{howpublished}{\url{https://github.com/andreaskipf/learnedcardinalities/blob/master/workloads/job-light.sql}}.
\newblock


\bibitem[\protect\citeauthoryear{??}{msc}{tory}]%
        {mscn_codes_url}
 \bibinfo{year}{MSCN Github repository}\natexlab{}.
\newblock \bibinfo{howpublished}{\url{https://github.com/andreaskipf/learnedcardinalities}}.
\newblock


\bibitem[\protect\citeauthoryear{??}{neu}{tory}]%
        {neurocard_codes_url}
 \bibinfo{year}{NeuroCard Github repository}\natexlab{}.
\newblock \bibinfo{howpublished}{\url{https://github.com/neurocard/neurocard}}.
\newblock


\bibitem[\protect\citeauthoryear{??}{tpc}{line}]%
        {tpch_url}
 \bibinfo{year}{online}\natexlab{}.
\newblock \bibinfo{howpublished}{\url{https://www.tpc.org/tpc_documents_current_versions/current_specifications5.asp}}.
\newblock


\bibitem[\protect\citeauthoryear{??}{sta}{line}]%
        {stats_web_url}
 \bibinfo{year}{online}\natexlab{}.
\newblock \bibinfo{howpublished}{\url{https://stats.stackexchange.com/}}.
\newblock


\bibitem[\protect\citeauthoryear{??}{pri}{tory}]%
        {price_codes_url}
 \bibinfo{year}{PRICE Github repository}\natexlab{}.
\newblock \bibinfo{howpublished}{\url{https://github.com/StCarmen/PRICE}}.
\newblock


\bibitem[\protect\citeauthoryear{??}{STA}{aset}]%
        {STATS_url}
 \bibinfo{year}{STATS Dataset}\natexlab{}.
\newblock \bibinfo{howpublished}{\url{https://relational.fit.cvut.cz/dataset/Stats}}.
\newblock


\bibitem[\protect\citeauthoryear{??}{gen}{aset}]%
        {genome_url}
 \bibinfo{year}{VisualGenome Dataset}\natexlab{}.
\newblock \bibinfo{howpublished}{\url{https://relational-data.org/dataset/VisualGenome}}.
\newblock


\bibitem[\protect\citeauthoryear{Agnihotri, Koldehofe, Binnig, and Luthra}{Agnihotri et~al\mbox{.}}{2023}]%
        {agnihotri_zero-shot_2023}
\bibfield{author}{\bibinfo{person}{Pratyush Agnihotri}, \bibinfo{person}{Boris Koldehofe}, \bibinfo{person}{Carsten Binnig}, {and} \bibinfo{person}{Manisha Luthra}.} \bibinfo{year}{2023}\natexlab{}.
\newblock \showarticletitle{Zero-{Shot} {Cost} {Models} for {Parallel} {Stream} {Processing}}. In \bibinfo{booktitle}{\emph{Proceedings of the {Sixth} {International} {Workshop} on {Exploiting} {Artificial} {Intelligence} {Techniques} for {Data} {Management}}}. \bibinfo{publisher}{ACM}, \bibinfo{address}{Seattle WA USA}, \bibinfo{pages}{1--5}.
\newblock
\showISBNx{9798400701931}
\urldef\tempurl%
\url{https://doi.org/10.1145/3593078.3593934}
\showDOI{\tempurl}


\bibitem[\protect\citeauthoryear{Brown, Mann, Ryder, Subbiah, Kaplan, Dhariwal, Neelakantan, Shyam, Sastry, Askell, Agarwal, Herbert-Voss, Krueger, Henighan, Child, Ramesh, Ziegler, Wu, Winter, Hesse, Chen, Sigler, Litwin, Gray, Chess, Clark, Berner, McCandlish, Radford, Sutskever, and Amodei}{Brown et~al\mbox{.}}{2020}]%
        {10.5555/3495724.3495883}
\bibfield{author}{\bibinfo{person}{Tom~B. Brown}, \bibinfo{person}{Benjamin Mann}, \bibinfo{person}{Nick Ryder}, \bibinfo{person}{Melanie Subbiah}, \bibinfo{person}{Jared Kaplan}, \bibinfo{person}{Prafulla Dhariwal}, \bibinfo{person}{Arvind Neelakantan}, \bibinfo{person}{Pranav Shyam}, \bibinfo{person}{Girish Sastry}, \bibinfo{person}{Amanda Askell}, \bibinfo{person}{Sandhini Agarwal}, \bibinfo{person}{Ariel Herbert-Voss}, \bibinfo{person}{Gretchen Krueger}, \bibinfo{person}{Tom Henighan}, \bibinfo{person}{Rewon Child}, \bibinfo{person}{Aditya Ramesh}, \bibinfo{person}{Daniel~M. Ziegler}, \bibinfo{person}{Jeffrey Wu}, \bibinfo{person}{Clemens Winter}, \bibinfo{person}{Christopher Hesse}, \bibinfo{person}{Mark Chen}, \bibinfo{person}{Eric Sigler}, \bibinfo{person}{Mateusz Litwin}, \bibinfo{person}{Scott Gray}, \bibinfo{person}{Benjamin Chess}, \bibinfo{person}{Jack Clark}, \bibinfo{person}{Christopher Berner}, \bibinfo{person}{Sam McCandlish}, \bibinfo{person}{Alec Radford}, \bibinfo{person}{Ilya Sutskever},
  {and} \bibinfo{person}{Dario Amodei}.} \bibinfo{year}{2020}\natexlab{}.
\newblock \showarticletitle{Language models are few-shot learners}. In \bibinfo{booktitle}{\emph{Proceedings of the 34th International Conference on Neural Information Processing Systems}} \emph{(\bibinfo{series}{NIPS '20})}. \bibinfo{publisher}{Curran Associates Inc.}, \bibinfo{address}{Red Hook, NY, USA}, Article \bibinfo{articleno}{159}, \bibinfo{numpages}{25}~pages.
\newblock
\showISBNx{9781713829546}


\bibitem[\protect\citeauthoryear{Bruno, Chaudhuri, and Gravano}{Bruno et~al\mbox{.}}{2001}]%
        {DBLP:conf/sigmod/BrunoCG01}
\bibfield{author}{\bibinfo{person}{Nicolas Bruno}, \bibinfo{person}{Surajit Chaudhuri}, {and} \bibinfo{person}{Luis Gravano}.} \bibinfo{year}{2001}\natexlab{}.
\newblock \showarticletitle{STHoles: {A} Multidimensional Workload-Aware Histogram}. In \bibinfo{booktitle}{\emph{SIGMOD}}. \bibinfo{pages}{211--222}.
\newblock


\bibitem[\protect\citeauthoryear{Chow and Liu}{Chow and Liu}{1968}]%
        {DBLP:journals/tit/ChowL68}
\bibfield{author}{\bibinfo{person}{C.~K. Chow} {and} \bibinfo{person}{C.~N. Liu}.} \bibinfo{year}{1968}\natexlab{}.
\newblock \showarticletitle{Approximating discrete probability distributions with dependence trees}.
\newblock \bibinfo{journal}{\emph{{IEEE} Trans. Inf. Theory}} \bibinfo{volume}{14}, \bibinfo{number}{3} (\bibinfo{year}{1968}), \bibinfo{pages}{462--467}.
\newblock


\bibitem[\protect\citeauthoryear{Deshpande, Garofalakis, and Rastogi}{Deshpande et~al\mbox{.}}{2001}]%
        {deshpande2001independence}
\bibfield{author}{\bibinfo{person}{Amol Deshpande}, \bibinfo{person}{Minos Garofalakis}, {and} \bibinfo{person}{Rajeev Rastogi}.} \bibinfo{year}{2001}\natexlab{}.
\newblock \showarticletitle{{Independence is good: Dependency-based histogram synopses for high-dimensional data}}.
\newblock \bibinfo{journal}{\emph{ACM SIGMOD Record}} \bibinfo{volume}{30}, \bibinfo{number}{2} (\bibinfo{year}{2001}), \bibinfo{pages}{199--210}.
\newblock


\bibitem[\protect\citeauthoryear{Devlin, Chang, Lee, and Toutanova}{Devlin et~al\mbox{.}}{2019}]%
        {devlin-etal-2019-bert}
\bibfield{author}{\bibinfo{person}{Jacob Devlin}, \bibinfo{person}{Ming-Wei Chang}, \bibinfo{person}{Kenton Lee}, {and} \bibinfo{person}{Kristina Toutanova}.} \bibinfo{year}{2019}\natexlab{}.
\newblock \showarticletitle{{BERT}: Pre-training of Deep Bidirectional Transformers for Language Understanding}. In \bibinfo{booktitle}{\emph{Proceedings of the 2019 Conference of the North {A}merican Chapter of the Association for Computational Linguistics: Human Language Technologies, Volume 1 (Long and Short Papers)}}, \bibfield{editor}{\bibinfo{person}{Jill Burstein}, \bibinfo{person}{Christy Doran}, {and} \bibinfo{person}{Thamar Solorio}} (Eds.). \bibinfo{publisher}{Association for Computational Linguistics}, \bibinfo{address}{Minneapolis, Minnesota}, \bibinfo{pages}{4171--4186}.
\newblock
\urldef\tempurl%
\url{https://doi.org/10.18653/v1/N19-1423}
\showDOI{\tempurl}


\bibitem[\protect\citeauthoryear{Documentation~12}{Documentation~12}{2020}]%
        {psql2020}
\bibfield{author}{\bibinfo{person}{Postgresql Documentation~12}.} \bibinfo{year}{2020}\natexlab{}.
\newblock \showarticletitle{Chapter 70.1. Row Estimation Examples}.
\newblock \bibinfo{journal}{\emph{https://www.postgresql.org/docs/current/row-estimation-examples.html}} (\bibinfo{year}{2020}).
\newblock


\bibitem[\protect\citeauthoryear{Dutt, Wang, Narasayya, and Chaudhuri}{Dutt et~al\mbox{.}}{2020}]%
        {pvldb:DuttWNC20}
\bibfield{author}{\bibinfo{person}{Anshuman Dutt}, \bibinfo{person}{Chi Wang}, \bibinfo{person}{Vivek~R. Narasayya}, {and} \bibinfo{person}{Surajit Chaudhuri}.} \bibinfo{year}{2020}\natexlab{}.
\newblock \showarticletitle{Efficiently Approximating Selectivity Functions using Low Overhead Regression Models}.
\newblock \bibinfo{journal}{\emph{Proc. {VLDB} Endow.}} \bibinfo{volume}{13}, \bibinfo{number}{11} (\bibinfo{year}{2020}), \bibinfo{pages}{2215--2228}.
\newblock
\urldef\tempurl%
\url{http://www.vldb.org/pvldb/vol13/p2215-dutt.pdf}
\showURL{%
\tempurl}


\bibitem[\protect\citeauthoryear{Dutt, Wang, Nazi, Kandula, Narasayya, and Chaudhuri}{Dutt et~al\mbox{.}}{2019}]%
        {DBLP:journals/pvldb/DuttWNKNC19}
\bibfield{author}{\bibinfo{person}{Anshuman Dutt}, \bibinfo{person}{Chi Wang}, \bibinfo{person}{Azade Nazi}, \bibinfo{person}{Srikanth Kandula}, \bibinfo{person}{Vivek~R. Narasayya}, {and} \bibinfo{person}{Surajit Chaudhuri}.} \bibinfo{year}{2019}\natexlab{}.
\newblock \showarticletitle{Selectivity Estimation for Range Predicates using Lightweight Models}.
\newblock \bibinfo{journal}{\emph{Proc. {VLDB} Endow.}} \bibinfo{volume}{12}, \bibinfo{number}{9} (\bibinfo{year}{2019}), \bibinfo{pages}{1044--1057}.
\newblock


\bibitem[\protect\citeauthoryear{Fuchs, He, and Lee}{Fuchs et~al\mbox{.}}{2007}]%
        {DBLP:journals/isci/FuchsHL07}
\bibfield{author}{\bibinfo{person}{Dennis Fuchs}, \bibinfo{person}{Zhen He}, {and} \bibinfo{person}{Byung~Suk Lee}.} \bibinfo{year}{2007}\natexlab{}.
\newblock \showarticletitle{Compressed histograms with arbitrary bucket layouts for selectivity estimation}.
\newblock \bibinfo{journal}{\emph{Inf. Sci.}} \bibinfo{volume}{177}, \bibinfo{number}{3} (\bibinfo{year}{2007}), \bibinfo{pages}{680--702}.
\newblock


\bibitem[\protect\citeauthoryear{Getoor, Taskar, and Koller}{Getoor et~al\mbox{.}}{2001}]%
        {getoor2001selectivity}
\bibfield{author}{\bibinfo{person}{Lise Getoor}, \bibinfo{person}{Benjamin Taskar}, {and} \bibinfo{person}{Daphne Koller}.} \bibinfo{year}{2001}\natexlab{}.
\newblock \showarticletitle{Selectivity estimation using probabilistic models}. In \bibinfo{booktitle}{\emph{SIGMOD}}. \bibinfo{pages}{461--472}.
\newblock


\bibitem[\protect\citeauthoryear{Group}{Group}{1996}]%
        {postgresql}
\bibfield{author}{\bibinfo{person}{PostgreSQL Global~Development Group}.} \bibinfo{year}{1996}\natexlab{}.
\newblock \showarticletitle{PostgreSQL}.
\newblock \bibinfo{howpublished}{https://www.postgresql.org}.
\newblock  (\bibinfo{year}{1996}).
\newblock
\newblock
\shownote{Accessed: 2022-10-28.}


\bibitem[\protect\citeauthoryear{Gunopulos, Kollios, Tsotras, and Domeniconi}{Gunopulos et~al\mbox{.}}{2000}]%
        {gunopulos2000approximating}
\bibfield{author}{\bibinfo{person}{Dimitrios Gunopulos}, \bibinfo{person}{George Kollios}, \bibinfo{person}{Vassilis~J Tsotras}, {and} \bibinfo{person}{Carlotta Domeniconi}.} \bibinfo{year}{2000}\natexlab{}.
\newblock \showarticletitle{{Approximating multi-dimensional aggregate range queries over real attributes}}. In \bibinfo{booktitle}{\emph{SIGMOD}}. \bibinfo{pages}{463--474}.
\newblock


\bibitem[\protect\citeauthoryear{Gunopulos, Kollios, Tsotras, and Domeniconi}{Gunopulos et~al\mbox{.}}{2005}]%
        {gunopulos2005selectivity}
\bibfield{author}{\bibinfo{person}{Dimitrios Gunopulos}, \bibinfo{person}{George Kollios}, \bibinfo{person}{Vassilis~J Tsotras}, {and} \bibinfo{person}{Carlotta Domeniconi}.} \bibinfo{year}{2005}\natexlab{}.
\newblock \showarticletitle{Selectivity estimators for multidimensional range queries over real attributes}.
\newblock \bibinfo{journal}{\emph{The VLDB Journal}} \bibinfo{volume}{14}, \bibinfo{number}{2} (\bibinfo{year}{2005}), \bibinfo{pages}{137--154}.
\newblock


\bibitem[\protect\citeauthoryear{Han, Wu, Wu, Zhu, Yang, Tan, Zeng, Cong, Qin, Pfadler, Qian, Zhou, Li, and Cui}{Han et~al\mbox{.}}{2021}]%
        {DBLP:journals/pvldb/HanWWZYTZCQPQZL21}
\bibfield{author}{\bibinfo{person}{Yuxing Han}, \bibinfo{person}{Ziniu Wu}, \bibinfo{person}{Peizhi Wu}, \bibinfo{person}{Rong Zhu}, \bibinfo{person}{Jingyi Yang}, \bibinfo{person}{Liang~Wei Tan}, \bibinfo{person}{Kai Zeng}, \bibinfo{person}{Gao Cong}, \bibinfo{person}{Yanzhao Qin}, \bibinfo{person}{Andreas Pfadler}, \bibinfo{person}{Zhengping Qian}, \bibinfo{person}{Jingren Zhou}, \bibinfo{person}{Jiangneng Li}, {and} \bibinfo{person}{Bin Cui}.} \bibinfo{year}{2021}\natexlab{}.
\newblock \showarticletitle{Cardinality Estimation in {DBMS:} {A} Comprehensive Benchmark Evaluation}.
\newblock \bibinfo{journal}{\emph{Proc. {VLDB} Endow.}} \bibinfo{volume}{15}, \bibinfo{number}{4} (\bibinfo{year}{2021}), \bibinfo{pages}{752--765}.
\newblock


\bibitem[\protect\citeauthoryear{Hasan, Thirumuruganathan, Augustine, Koudas, and Das}{Hasan et~al\mbox{.}}{2019}]%
        {hasan2019multi}
\bibfield{author}{\bibinfo{person}{Shohedul Hasan}, \bibinfo{person}{Saravanan Thirumuruganathan}, \bibinfo{person}{Jees Augustine}, \bibinfo{person}{Nick Koudas}, {and} \bibinfo{person}{Gautam Das}.} \bibinfo{year}{2019}\natexlab{}.
\newblock \showarticletitle{Multi-attribute selectivity estimation using deep learning}. In \bibinfo{booktitle}{\emph{SIGMOD}}.
\newblock


\bibitem[\protect\citeauthoryear{Hastie, Tibshirani, and Friedman}{Hastie et~al\mbox{.}}{2009}]%
        {DBLP:books/lib/HastieTF09}
\bibfield{author}{\bibinfo{person}{Trevor Hastie}, \bibinfo{person}{Robert Tibshirani}, {and} \bibinfo{person}{Jerome~H. Friedman}.} \bibinfo{year}{2009}\natexlab{}.
\newblock \bibinfo{booktitle}{\emph{The Elements of Statistical Learning: Data Mining, Inference, and Prediction, 2nd Edition}}.
\newblock \bibinfo{publisher}{Springer}.
\newblock


\bibitem[\protect\citeauthoryear{Heimel, Kiefer, and Markl}{Heimel et~al\mbox{.}}{2015}]%
        {heimel2015self}
\bibfield{author}{\bibinfo{person}{Max Heimel}, \bibinfo{person}{Martin Kiefer}, {and} \bibinfo{person}{Volker Markl}.} \bibinfo{year}{2015}\natexlab{}.
\newblock \showarticletitle{Self-tuning, gpu-accelerated kernel density models for multidimensional selectivity estimation}. In \bibinfo{booktitle}{\emph{SIGMOD}}. \bibinfo{pages}{1477--1492}.
\newblock


\bibitem[\protect\citeauthoryear{Heinrich, Luthra, Kornmayer, and Binnig}{Heinrich et~al\mbox{.}}{2022}]%
        {heinrich_zero-shot_2022}
\bibfield{author}{\bibinfo{person}{Roman Heinrich}, \bibinfo{person}{Manisha Luthra}, \bibinfo{person}{Harald Kornmayer}, {and} \bibinfo{person}{Carsten Binnig}.} \bibinfo{year}{2022}\natexlab{}.
\newblock \showarticletitle{Zero-shot cost models for distributed stream processing}. In \bibinfo{booktitle}{\emph{Proceedings of the 16th {ACM} {International} {Conference} on {Distributed} and {Event}-{Based} {Systems}}}. \bibinfo{publisher}{ACM}, \bibinfo{address}{Copenhagen Denmark}, \bibinfo{pages}{85--90}.
\newblock
\showISBNx{978-1-4503-9308-9}
\urldef\tempurl%
\url{https://doi.org/10.1145/3524860.3539639}
\showDOI{\tempurl}


\bibitem[\protect\citeauthoryear{Hilprecht and Binnig}{Hilprecht and Binnig}{2022a}]%
        {hilprecht_one_2022}
\bibfield{author}{\bibinfo{person}{Benjamin Hilprecht} {and} \bibinfo{person}{Carsten Binnig}.} \bibinfo{year}{2022}\natexlab{a}.
\newblock \bibinfo{title}{One {Model} to {Rule} them {All}: {Towards} {Zero}-{Shot} {Learning} for {Databases}}.
\newblock
\newblock
\urldef\tempurl%
\url{http://arxiv.org/abs/2105.00642}
\showURL{%
\tempurl}
\newblock
\shownote{arXiv:2105.00642 [cs].}


\bibitem[\protect\citeauthoryear{Hilprecht and Binnig}{Hilprecht and Binnig}{2022b}]%
        {hilprecht_zero-shot_2022}
\bibfield{author}{\bibinfo{person}{Benjamin Hilprecht} {and} \bibinfo{person}{Carsten Binnig}.} \bibinfo{year}{2022}\natexlab{b}.
\newblock \showarticletitle{Zero-shot cost models for out-of-the-box learned cost prediction}.
\newblock \bibinfo{journal}{\emph{Proceedings of the VLDB Endowment}} \bibinfo{volume}{15}, \bibinfo{number}{11} (\bibinfo{date}{July} \bibinfo{year}{2022}), \bibinfo{pages}{2361--2374}.
\newblock
\showISSN{2150-8097}
\urldef\tempurl%
\url{https://doi.org/10.14778/3551793.3551799}
\showDOI{\tempurl}


\bibitem[\protect\citeauthoryear{Hilprecht, Schmidt, Kulessa, Molina, Kersting, and Binnig}{Hilprecht et~al\mbox{.}}{2020}]%
        {DBLP:journals/pvldb/HilprechtSKMKB20}
\bibfield{author}{\bibinfo{person}{Benjamin Hilprecht}, \bibinfo{person}{Andreas Schmidt}, \bibinfo{person}{Moritz Kulessa}, \bibinfo{person}{Alejandro Molina}, \bibinfo{person}{Kristian Kersting}, {and} \bibinfo{person}{Carsten Binnig}.} \bibinfo{year}{2020}\natexlab{}.
\newblock \showarticletitle{DeepDB: Learn from Data, not from Queries!}
\newblock \bibinfo{journal}{\emph{Proc. {VLDB} Endow.}} \bibinfo{volume}{13}, \bibinfo{number}{7} (\bibinfo{year}{2020}), \bibinfo{pages}{992--1005}.
\newblock


\bibitem[\protect\citeauthoryear{Khachatryan, M{\"{u}}ller, Stier, and B{\"{o}}hm}{Khachatryan et~al\mbox{.}}{2015}]%
        {DBLP:journals/tkde/KhachatryanMSB15}
\bibfield{author}{\bibinfo{person}{Andranik Khachatryan}, \bibinfo{person}{Emmanuel M{\"{u}}ller}, \bibinfo{person}{Christian Stier}, {and} \bibinfo{person}{Klemens B{\"{o}}hm}.} \bibinfo{year}{2015}\natexlab{}.
\newblock \showarticletitle{Improving Accuracy and Robustness of Self-Tuning Histograms by Subspace Clustering}.
\newblock \bibinfo{journal}{\emph{{IEEE} Trans. Knowl. Data Eng.}} \bibinfo{volume}{27}, \bibinfo{number}{9} (\bibinfo{year}{2015}), \bibinfo{pages}{2377--2389}.
\newblock


\bibitem[\protect\citeauthoryear{Kiefer, Heimel, Bre{\ss}, and Markl}{Kiefer et~al\mbox{.}}{2017}]%
        {DBLP:journals/pvldb/KieferHBM17}
\bibfield{author}{\bibinfo{person}{Martin Kiefer}, \bibinfo{person}{Max Heimel}, \bibinfo{person}{Sebastian Bre{\ss}}, {and} \bibinfo{person}{Volker Markl}.} \bibinfo{year}{2017}\natexlab{}.
\newblock \showarticletitle{Estimating Join Selectivities using Bandwidth-Optimized Kernel Density Models}.
\newblock \bibinfo{journal}{\emph{Proc. {VLDB} Endow.}} \bibinfo{volume}{10}, \bibinfo{number}{13} (\bibinfo{year}{2017}), \bibinfo{pages}{2085--2096}.
\newblock


\bibitem[\protect\citeauthoryear{Kim, Jung, Seo, Han, Choi, and Chong}{Kim et~al\mbox{.}}{2022}]%
        {kim_learned_2022}
\bibfield{author}{\bibinfo{person}{Kyoungmin Kim}, \bibinfo{person}{Jisung Jung}, \bibinfo{person}{In Seo}, \bibinfo{person}{Wook-Shin Han}, \bibinfo{person}{Kangwoo Choi}, {and} \bibinfo{person}{Jaehyok Chong}.} \bibinfo{year}{2022}\natexlab{}.
\newblock \showarticletitle{Learned {Cardinality} {Estimation}: {An} {In}-depth {Study}}. In \bibinfo{booktitle}{\emph{Proceedings of the 2022 {International} {Conference} on {Management} of {Data}}}. \bibinfo{publisher}{ACM}, \bibinfo{address}{Philadelphia PA USA}, \bibinfo{pages}{1214--1227}.
\newblock
\showISBNx{978-1-4503-9249-5}
\urldef\tempurl%
\url{https://doi.org/10.1145/3514221.3526154}
\showDOI{\tempurl}


\bibitem[\protect\citeauthoryear{Kingma and Ba}{Kingma and Ba}{2015}]%
        {DBLP:journals/corr/KingmaB14}
\bibfield{author}{\bibinfo{person}{Diederik~P. Kingma} {and} \bibinfo{person}{Jimmy Ba}.} \bibinfo{year}{2015}\natexlab{}.
\newblock \showarticletitle{Adam: {A} Method for Stochastic Optimization}. In \bibinfo{booktitle}{\emph{ICLR}}.
\newblock


\bibitem[\protect\citeauthoryear{Kipf, Kipf, Radke, Leis, Boncz, and Kemper}{Kipf et~al\mbox{.}}{2019}]%
        {DBLP:conf/cidr/KipfKRLBK19}
\bibfield{author}{\bibinfo{person}{Andreas Kipf}, \bibinfo{person}{Thomas Kipf}, \bibinfo{person}{Bernhard Radke}, \bibinfo{person}{Viktor Leis}, \bibinfo{person}{Peter~A. Boncz}, {and} \bibinfo{person}{Alfons Kemper}.} \bibinfo{year}{2019}\natexlab{}.
\newblock \showarticletitle{Learned Cardinalities: Estimating Correlated Joins with Deep Learning}. In \bibinfo{booktitle}{\emph{CIDR}}.
\newblock


\bibitem[\protect\citeauthoryear{Lakshminarayanan, Pritzel, and Blundell}{Lakshminarayanan et~al\mbox{.}}{2017}]%
        {DBLP:conf/nips/Lakshminarayanan17}
\bibfield{author}{\bibinfo{person}{Balaji Lakshminarayanan}, \bibinfo{person}{Alexander Pritzel}, {and} \bibinfo{person}{Charles Blundell}.} \bibinfo{year}{2017}\natexlab{}.
\newblock \showarticletitle{Simple and Scalable Predictive Uncertainty Estimation using Deep Ensembles}. In \bibinfo{booktitle}{\emph{NIPS}}. \bibinfo{pages}{6402--6413}.
\newblock


\bibitem[\protect\citeauthoryear{Lee, Bahri, Novak, Schoenholz, Pennington, and Sohl{-}Dickstein}{Lee et~al\mbox{.}}{2018}]%
        {DBLP:conf/iclr/LeeBNSPS18}
\bibfield{author}{\bibinfo{person}{Jaehoon Lee}, \bibinfo{person}{Yasaman Bahri}, \bibinfo{person}{Roman Novak}, \bibinfo{person}{Samuel~S. Schoenholz}, \bibinfo{person}{Jeffrey Pennington}, {and} \bibinfo{person}{Jascha Sohl{-}Dickstein}.} \bibinfo{year}{2018}\natexlab{}.
\newblock \showarticletitle{Deep Neural Networks as Gaussian Processes}. In \bibinfo{booktitle}{\emph{ICLR}}.
\newblock


\bibitem[\protect\citeauthoryear{Leis, Gubichev, Mirchev, Boncz, Kemper, and Neumann}{Leis et~al\mbox{.}}{2015}]%
        {leis2015good}
\bibfield{author}{\bibinfo{person}{Viktor Leis}, \bibinfo{person}{Andrey Gubichev}, \bibinfo{person}{Atanas Mirchev}, \bibinfo{person}{Peter Boncz}, \bibinfo{person}{Alfons Kemper}, {and} \bibinfo{person}{Thomas Neumann}.} \bibinfo{year}{2015}\natexlab{}.
\newblock \showarticletitle{How good are query optimizers, really?}
\newblock \bibinfo{journal}{\emph{PVLDB}} \bibinfo{volume}{9}, \bibinfo{number}{3} (\bibinfo{year}{2015}), \bibinfo{pages}{204--215}.
\newblock


\bibitem[\protect\citeauthoryear{Leis, Radke, Gubichev, Kemper, and Neumann}{Leis et~al\mbox{.}}{2017}]%
        {leis2017cardinality}
\bibfield{author}{\bibinfo{person}{Viktor Leis}, \bibinfo{person}{Bernhard Radke}, \bibinfo{person}{Andrey Gubichev}, \bibinfo{person}{Alfons Kemper}, {and} \bibinfo{person}{Thomas Neumann}.} \bibinfo{year}{2017}\natexlab{}.
\newblock \showarticletitle{Cardinality Estimation Done Right: Index-Based Join Sampling}. In \bibinfo{booktitle}{\emph{CIDR}}.
\newblock


\bibitem[\protect\citeauthoryear{Li, Wu, Yi, and Zhao}{Li et~al\mbox{.}}{2016}]%
        {li2016wander}
\bibfield{author}{\bibinfo{person}{Feifei Li}, \bibinfo{person}{Bin Wu}, \bibinfo{person}{Ke Yi}, {and} \bibinfo{person}{Zhuoyue Zhao}.} \bibinfo{year}{2016}\natexlab{}.
\newblock \showarticletitle{Wander join: Online aggregation via random walks}. In \bibinfo{booktitle}{\emph{SIGMOD}}. \bibinfo{pages}{615--629}.
\newblock


\bibitem[\protect\citeauthoryear{Li, Wei, Zhu, Ding, Zhou, and Lu}{Li et~al\mbox{.}}{2023}]%
        {li_alece_2023}
\bibfield{author}{\bibinfo{person}{Pengfei Li}, \bibinfo{person}{Wenqing Wei}, \bibinfo{person}{Rong Zhu}, \bibinfo{person}{Bolin Ding}, \bibinfo{person}{Jingren Zhou}, {and} \bibinfo{person}{Hua Lu}.} \bibinfo{year}{2023}\natexlab{}.
\newblock \showarticletitle{{ALECE}: {An} {Attention}-based {Learned} {Cardinality} {Estimator} for {SPJ} {Queries} on {Dynamic} {Workloads}}.
\newblock \bibinfo{journal}{\emph{Proceedings of the VLDB Endowment}} \bibinfo{volume}{17}, \bibinfo{number}{2} (\bibinfo{date}{Oct.} \bibinfo{year}{2023}), \bibinfo{pages}{197--210}.
\newblock
\showISSN{2150-8097}
\urldef\tempurl%
\url{https://doi.org/10.14778/3626292.3626302}
\showDOI{\tempurl}


\bibitem[\protect\citeauthoryear{Liu, Xu, Yu, Corvinelli, and Zuzarte}{Liu et~al\mbox{.}}{2015}]%
        {cascon:LiuXYCZ15}
\bibfield{author}{\bibinfo{person}{Henry Liu}, \bibinfo{person}{Mingbin Xu}, \bibinfo{person}{Ziting Yu}, \bibinfo{person}{Vincent Corvinelli}, {and} \bibinfo{person}{Calisto Zuzarte}.} \bibinfo{year}{2015}\natexlab{}.
\newblock \showarticletitle{Cardinality estimation using neural networks}. In \bibinfo{booktitle}{\emph{Proceedings of 25th Annual International Conference on Computer Science and Software Engineering, {CASCON} 2015, Markham, Ontario, Canada, 2-4 November, 2015}}. \bibinfo{publisher}{{IBM} / {ACM}}, \bibinfo{pages}{53--59}.
\newblock
\urldef\tempurl%
\url{http://dl.acm.org/citation.cfm?id=2886453}
\showURL{%
\tempurl}


\bibitem[\protect\citeauthoryear{Liu, Dong, Li, and Zhou}{Liu et~al\mbox{.}}{2021}]%
        {DBLP:journals/pvldb/LiuD0Z21}
\bibfield{author}{\bibinfo{person}{Jie Liu}, \bibinfo{person}{Wenqian Dong}, \bibinfo{person}{Dong Li}, {and} \bibinfo{person}{Qingqing Zhou}.} \bibinfo{year}{2021}\natexlab{}.
\newblock \showarticletitle{Fauce: Fast and Accurate Deep Ensembles with Uncertainty for Cardinality Estimation}.
\newblock \bibinfo{journal}{\emph{Proc. {VLDB} Endow.}} \bibinfo{volume}{14}, \bibinfo{number}{11} (\bibinfo{year}{2021}), \bibinfo{pages}{1950--1963}.
\newblock


\bibitem[\protect\citeauthoryear{Liu, Ott, Goyal, Du, Joshi, Chen, Levy, Lewis, Zettlemoyer, and Stoyanov}{Liu et~al\mbox{.}}{2019}]%
        {roberta_liu2019roberta}
\bibfield{author}{\bibinfo{person}{Yinhan Liu}, \bibinfo{person}{Myle Ott}, \bibinfo{person}{Naman Goyal}, \bibinfo{person}{Jingfei Du}, \bibinfo{person}{Mandar Joshi}, \bibinfo{person}{Danqi Chen}, \bibinfo{person}{Omer Levy}, \bibinfo{person}{Mike Lewis}, \bibinfo{person}{Luke Zettlemoyer}, {and} \bibinfo{person}{Veselin Stoyanov}.} \bibinfo{year}{2019}\natexlab{}.
\newblock \showarticletitle{Roberta: A robustly optimized bert pretraining approach}.
\newblock \bibinfo{journal}{\emph{arXiv preprint arXiv:1907.11692}} (\bibinfo{year}{2019}).
\newblock


\bibitem[\protect\citeauthoryear{Lopes, Guyer, and Gene}{Lopes et~al\mbox{.}}{2019}]%
        {sqlserver2019}
\bibfield{author}{\bibinfo{person}{Pedro Lopes}, \bibinfo{person}{Craig Guyer}, {and} \bibinfo{person}{Milener Gene}.} \bibinfo{year}{2019}\natexlab{}.
\newblock \showarticletitle{Sql docs: cardinality estimation (SQL Server)}.
\newblock \bibinfo{journal}{\emph{https://docs.microsoft.com/en-us/sql/relational-databases/performance/cardinality-estimation-sql-server?view=sql-server-ver15}} (\bibinfo{year}{2019}).
\newblock


\bibitem[\protect\citeauthoryear{Lopez-Paz, Hennig, and Schölkopf}{Lopez-Paz et~al\mbox{.}}{[n.d.]}]%
        {lopez-paz_randomized_nodate}
\bibfield{author}{\bibinfo{person}{David Lopez-Paz}, \bibinfo{person}{Philipp Hennig}, {and} \bibinfo{person}{Bernhard Schölkopf}.} \bibinfo{year}{[n.d.]}\natexlab{}.
\newblock \showarticletitle{The {Randomized} {Dependence} {Coefficient}}.
\newblock  (\bibinfo{year}{[n.\,d.]}).
\newblock


\bibitem[\protect\citeauthoryear{Lu, Kandula, König, and Chaudhuri}{Lu et~al\mbox{.}}{2021}]%
        {lu_pre-training_2021}
\bibfield{author}{\bibinfo{person}{Yao Lu}, \bibinfo{person}{Srikanth Kandula}, \bibinfo{person}{Arnd~Christian König}, {and} \bibinfo{person}{Surajit Chaudhuri}.} \bibinfo{year}{2021}\natexlab{}.
\newblock \showarticletitle{Pre-training summarization models of structured datasets for cardinality estimation}.
\newblock \bibinfo{journal}{\emph{Proceedings of the VLDB Endowment}} \bibinfo{volume}{15}, \bibinfo{number}{3} (\bibinfo{date}{Nov.} \bibinfo{year}{2021}), \bibinfo{pages}{414--426}.
\newblock
\showISSN{2150-8097}
\urldef\tempurl%
\url{https://doi.org/10.14778/3494124.3494127}
\showDOI{\tempurl}


\bibitem[\protect\citeauthoryear{{MariaDB Server Documentation}}{{MariaDB Server Documentation}}{2020}]%
        {MariaDB2020Statistics}
\bibfield{author}{\bibinfo{person}{{MariaDB Server Documentation}}.} \bibinfo{year}{2020}\natexlab{}.
\newblock \bibinfo{title}{Statistics for optimizing queries: InnoDB persistent statistics}.
\newblock \bibinfo{howpublished}{\url{https://mariadb.com/kb/en/innodb-persistent-statistics/}}.
\newblock
\newblock
\shownote{Accessed: 2020.}


\bibitem[\protect\citeauthoryear{Metwally, Agrawal, and Abbadi}{Metwally et~al\mbox{.}}{2006}]%
        {10.1145/1166074.1166084}
\bibfield{author}{\bibinfo{person}{Ahmed Metwally}, \bibinfo{person}{Divyakant Agrawal}, {and} \bibinfo{person}{Amr~El Abbadi}.} \bibinfo{year}{2006}\natexlab{}.
\newblock \showarticletitle{An integrated efficient solution for computing frequent and top-k elements in data streams}.
\newblock \bibinfo{journal}{\emph{ACM Trans. Database Syst.}} \bibinfo{volume}{31}, \bibinfo{number}{3} (\bibinfo{date}{sep} \bibinfo{year}{2006}), \bibinfo{pages}{1095–1133}.
\newblock
\showISSN{0362-5915}
\urldef\tempurl%
\url{https://doi.org/10.1145/1166074.1166084}
\showDOI{\tempurl}


\bibitem[\protect\citeauthoryear{Moerkotte, Neumann, and Steidl}{Moerkotte et~al\mbox{.}}{2009}]%
        {DBLP:journals/pvldb/MoerkotteNS09}
\bibfield{author}{\bibinfo{person}{Guido Moerkotte}, \bibinfo{person}{Thomas Neumann}, {and} \bibinfo{person}{Gabriele Steidl}.} \bibinfo{year}{2009}\natexlab{}.
\newblock \showarticletitle{Preventing Bad Plans by Bounding the Impact of Cardinality Estimation Errors}.
\newblock \bibinfo{journal}{\emph{Proc. {VLDB} Endow.}} \bibinfo{volume}{2}, \bibinfo{number}{1} (\bibinfo{year}{2009}), \bibinfo{pages}{982--993}.
\newblock


\bibitem[\protect\citeauthoryear{Motl and Schulte}{Motl and Schulte}{2015}]%
        {DBLP:journals/corr/MotlS15}
\bibfield{author}{\bibinfo{person}{Jan Motl} {and} \bibinfo{person}{Oliver Schulte}.} \bibinfo{year}{2015}\natexlab{}.
\newblock \showarticletitle{The {CTU} Prague Relational Learning Repository}.
\newblock \bibinfo{journal}{\emph{CoRR}}  \bibinfo{volume}{abs/1511.03086} (\bibinfo{year}{2015}).
\newblock
\showeprint[arXiv]{1511.03086}
\urldef\tempurl%
\url{http://arxiv.org/abs/1511.03086}
\showURL{%
\tempurl}


\bibitem[\protect\citeauthoryear{Muralikrishna and DeWitt}{Muralikrishna and DeWitt}{1988}]%
        {muralikrishna1988equi}
\bibfield{author}{\bibinfo{person}{M Muralikrishna} {and} \bibinfo{person}{David~J DeWitt}.} \bibinfo{year}{1988}\natexlab{}.
\newblock \showarticletitle{Equi-depth multidimensional histograms}. In \bibinfo{booktitle}{\emph{Proceedings of the 1988 ACM SIGMOD international conference on Management of data}}. \bibinfo{pages}{28--36}.
\newblock


\bibitem[\protect\citeauthoryear{Negi, Wu, Kipf, Tatbul, Marcus, Madden, Kraska, and Alizadeh}{Negi et~al\mbox{.}}{2023}]%
        {DBLP:journals/pvldb/NegiWKTMMKA23}
\bibfield{author}{\bibinfo{person}{Parimarjan Negi}, \bibinfo{person}{Ziniu Wu}, \bibinfo{person}{Andreas Kipf}, \bibinfo{person}{Nesime Tatbul}, \bibinfo{person}{Ryan Marcus}, \bibinfo{person}{Sam Madden}, \bibinfo{person}{Tim Kraska}, {and} \bibinfo{person}{Mohammad Alizadeh}.} \bibinfo{year}{2023}\natexlab{}.
\newblock \showarticletitle{Robust Query Driven Cardinality Estimation under Changing Workloads}.
\newblock \bibinfo{journal}{\emph{Proc. {VLDB} Endow.}} \bibinfo{volume}{16}, \bibinfo{number}{6} (\bibinfo{year}{2023}), \bibinfo{pages}{1520--1533}.
\newblock


\bibitem[\protect\citeauthoryear{O'Neil, O'Neil, Chen, and Revilak}{O'Neil et~al\mbox{.}}{2009}]%
        {10.1007/978-3-642-10424-4_17}
\bibfield{author}{\bibinfo{person}{Patrick O'Neil}, \bibinfo{person}{Elizabeth O'Neil}, \bibinfo{person}{Xuedong Chen}, {and} \bibinfo{person}{Stephen Revilak}.} \bibinfo{year}{2009}\natexlab{}.
\newblock \showarticletitle{The Star Schema Benchmark and Augmented Fact Table Indexing}. In \bibinfo{booktitle}{\emph{Performance Evaluation and Benchmarking}}, \bibfield{editor}{\bibinfo{person}{Raghunath Nambiar} {and} \bibinfo{person}{Meikel Poess}} (Eds.). \bibinfo{publisher}{Springer Berlin Heidelberg}, \bibinfo{address}{Berlin, Heidelberg}, \bibinfo{pages}{237--252}.
\newblock
\showISBNx{978-3-642-10424-4}


\bibitem[\protect\citeauthoryear{Paper}{Paper}{2019}]%
        {oracle2019}
\bibfield{author}{\bibinfo{person}{Oracle~White Paper}.} \bibinfo{year}{2019}\natexlab{}.
\newblock \showarticletitle{The optimizer In Oracle database 19c}.
\newblock \bibinfo{journal}{\emph{https://www.oracle.com/technetwork/database/bi-datawarehousing/twp-optimizer-with-oracledb-19c-5324206.pdf}} (\bibinfo{year}{2019}).
\newblock


\bibitem[\protect\citeauthoryear{Park, Zhong, and Mozafari}{Park et~al\mbox{.}}{2020}]%
        {sigmod:ParkZM20}
\bibfield{author}{\bibinfo{person}{Yongjoo Park}, \bibinfo{person}{Shucheng Zhong}, {and} \bibinfo{person}{Barzan Mozafari}.} \bibinfo{year}{2020}\natexlab{}.
\newblock \showarticletitle{QuickSel: Quick Selectivity Learning with Mixture Models}. In \bibinfo{booktitle}{\emph{Proceedings of the 2020 International Conference on Management of Data, {SIGMOD} Conference 2020, online conference [Portland, OR, USA], June 14-19, 2020}}. \bibinfo{publisher}{{ACM}}, \bibinfo{pages}{1017--1033}.
\newblock
\urldef\tempurl%
\url{https://doi.org/10.1145/3318464.3389727}
\showDOI{\tempurl}


\bibitem[\protect\citeauthoryear{Poon and Domingos}{Poon and Domingos}{2011}]%
        {DBLP:conf/uai/PoonD11}
\bibfield{author}{\bibinfo{person}{Hoifung Poon} {and} \bibinfo{person}{Pedro~M. Domingos}.} \bibinfo{year}{2011}\natexlab{}.
\newblock \showarticletitle{Sum-Product Networks: {A} New Deep Architecture}. In \bibinfo{booktitle}{\emph{UAI}}. \bibinfo{pages}{337--346}.
\newblock


\bibitem[\protect\citeauthoryear{Poosala and Ioannidis}{Poosala and Ioannidis}{1997}]%
        {poosala1997selectivity}
\bibfield{author}{\bibinfo{person}{Viswanath Poosala} {and} \bibinfo{person}{Yannis~E Ioannidis}.} \bibinfo{year}{1997}\natexlab{}.
\newblock \showarticletitle{Selectivity estimation without the attribute value independence assumption}. In \bibinfo{booktitle}{\emph{VLDB}}, Vol.~\bibinfo{volume}{97}. \bibinfo{pages}{486--495}.
\newblock


\bibitem[\protect\citeauthoryear{Reference~Manual}{Reference~Manual}{2020}]%
        {mysql2020}
\bibfield{author}{\bibinfo{person}{MySQL~8.0 Reference~Manual}.} \bibinfo{year}{2020}\natexlab{}.
\newblock \showarticletitle{Chapter 15.8.10.2 Configuring Non-Persistent Optimizer Statistics Parameters}.
\newblock \bibinfo{journal}{\emph{https://dev.mysql.com/doc/refman/8.0/en/innodb-statistics-estimation.html}} (\bibinfo{year}{2020}).
\newblock


\bibitem[\protect\citeauthoryear{Selinger, Astrahan, Chamberlin, Lorie, and Price}{Selinger et~al\mbox{.}}{1979}]%
        {selinger1979access}
\bibfield{author}{\bibinfo{person}{P~Griffiths Selinger}, \bibinfo{person}{Morton~M Astrahan}, \bibinfo{person}{Donald~D Chamberlin}, \bibinfo{person}{Raymond~A Lorie}, {and} \bibinfo{person}{Thomas~G Price}.} \bibinfo{year}{1979}\natexlab{}.
\newblock \showarticletitle{Access path selection in a relational database management system}. In \bibinfo{booktitle}{\emph{SIGMOD}}. \bibinfo{pages}{23--34}.
\newblock


\bibitem[\protect\citeauthoryear{Server~Documentation}{Server~Documentation}{2020}]%
        {mdb2020}
\bibfield{author}{\bibinfo{person}{MariaDB Server~Documentation}.} \bibinfo{year}{2020}\natexlab{}.
\newblock \showarticletitle{Statistics for optimizing queries: InnoDB persistent statistics}.
\newblock \bibinfo{journal}{\emph{https://mariadb.com/kb/en/innodb-persistent-statistics/}} (\bibinfo{year}{2020}).
\newblock


\bibitem[\protect\citeauthoryear{Srivastava, Haas, Markl, Kutsch, and Tran}{Srivastava et~al\mbox{.}}{2006}]%
        {DBLP:conf/icde/SrivastavaHMKT06}
\bibfield{author}{\bibinfo{person}{Utkarsh Srivastava}, \bibinfo{person}{Peter~J. Haas}, \bibinfo{person}{Volker Markl}, \bibinfo{person}{Marcel Kutsch}, {and} \bibinfo{person}{Tam~Minh Tran}.} \bibinfo{year}{2006}\natexlab{}.
\newblock \showarticletitle{{ISOMER:} Consistent Histogram Construction Using Query Feedback}. In \bibinfo{booktitle}{\emph{ICDE}}. \bibinfo{pages}{39}.
\newblock


\bibitem[\protect\citeauthoryear{Stillger, Lohman, Markl, and Kandil}{Stillger et~al\mbox{.}}{2001}]%
        {DBLP:conf/vldb/StillgerLMK01}
\bibfield{author}{\bibinfo{person}{Michael Stillger}, \bibinfo{person}{Guy~M. Lohman}, \bibinfo{person}{Volker Markl}, {and} \bibinfo{person}{Mokhtar Kandil}.} \bibinfo{year}{2001}\natexlab{}.
\newblock \showarticletitle{{LEO} - DB2's LEarning Optimizer}. In \bibinfo{booktitle}{\emph{VLDB}}. \bibinfo{pages}{19--28}.
\newblock


\bibitem[\protect\citeauthoryear{Sun, Zhang, Sun, Li, and Tang}{Sun et~al\mbox{.}}{2021}]%
        {DBLP:journals/pvldb/SunZSLT21}
\bibfield{author}{\bibinfo{person}{Ji Sun}, \bibinfo{person}{Jintao Zhang}, \bibinfo{person}{Zhaoyan Sun}, \bibinfo{person}{Guoliang Li}, {and} \bibinfo{person}{Nan Tang}.} \bibinfo{year}{2021}\natexlab{}.
\newblock \showarticletitle{Learned Cardinality Estimation: {A} Design Space Exploration and {A} Comparative Evaluation}.
\newblock \bibinfo{journal}{\emph{Proc. {VLDB} Endow.}} \bibinfo{volume}{15}, \bibinfo{number}{1} (\bibinfo{year}{2021}), \bibinfo{pages}{85--97}.
\newblock


\bibitem[\protect\citeauthoryear{Tzoumas, Deshpande, and Jensen}{Tzoumas et~al\mbox{.}}{2011}]%
        {tzoumas2011lightweight}
\bibfield{author}{\bibinfo{person}{Kostas Tzoumas}, \bibinfo{person}{Amol Deshpande}, {and} \bibinfo{person}{Christian~S Jensen}.} \bibinfo{year}{2011}\natexlab{}.
\newblock \showarticletitle{Lightweight graphical models for selectivity estimation without independence assumptions}.
\newblock \bibinfo{journal}{\emph{PVLDB}} \bibinfo{volume}{4}, \bibinfo{number}{11} (\bibinfo{year}{2011}), \bibinfo{pages}{852--863}.
\newblock


\bibitem[\protect\citeauthoryear{Vaswani, Shazeer, Parmar, Uszkoreit, Jones, Gomez, Kaiser, and Polosukhin}{Vaswani et~al\mbox{.}}{2017}]%
        {DBLP:conf/nips/VaswaniSPUJGKP17}
\bibfield{author}{\bibinfo{person}{Ashish Vaswani}, \bibinfo{person}{Noam Shazeer}, \bibinfo{person}{Niki Parmar}, \bibinfo{person}{Jakob Uszkoreit}, \bibinfo{person}{Llion Jones}, \bibinfo{person}{Aidan~N. Gomez}, \bibinfo{person}{Lukasz Kaiser}, {and} \bibinfo{person}{Illia Polosukhin}.} \bibinfo{year}{2017}\natexlab{}.
\newblock \showarticletitle{Attention is All you Need}. In \bibinfo{booktitle}{\emph{NeurIPS}}. \bibinfo{pages}{5998--6008}.
\newblock


\bibitem[\protect\citeauthoryear{Wang, Qu, Wu, Wang, and Zhou}{Wang et~al\mbox{.}}{2021}]%
        {wang2020ready}
\bibfield{author}{\bibinfo{person}{Xiaoying Wang}, \bibinfo{person}{Changbo Qu}, \bibinfo{person}{Weiyuan Wu}, \bibinfo{person}{Jiannan Wang}, {and} \bibinfo{person}{Qingqing Zhou}.} \bibinfo{year}{2021}\natexlab{}.
\newblock \showarticletitle{Are We Ready For Learned Cardinality Estimation?}
\newblock \bibinfo{journal}{\emph{VLDB}} \bibinfo{volume}{14}, \bibinfo{number}{9} (\bibinfo{year}{2021}), \bibinfo{pages}{1640--1654}.
\newblock


\bibitem[\protect\citeauthoryear{Weng, Zhu, Wu, Ding, Zheng, and Zhou}{Weng et~al\mbox{.}}{2024}]%
        {10.14778/3641204.3641205}
\bibfield{author}{\bibinfo{person}{Lianggui Weng}, \bibinfo{person}{Rong Zhu}, \bibinfo{person}{Di Wu}, \bibinfo{person}{Bolin Ding}, \bibinfo{person}{Bolong Zheng}, {and} \bibinfo{person}{Jingren Zhou}.} \bibinfo{year}{2024}\natexlab{}.
\newblock \showarticletitle{Eraser: Eliminating Performance Regression on Learned Query Optimizer}.
\newblock \bibinfo{journal}{\emph{Proc. VLDB Endow.}} \bibinfo{volume}{17}, \bibinfo{number}{5} (\bibinfo{date}{may} \bibinfo{year}{2024}), \bibinfo{pages}{926–938}.
\newblock
\showISSN{2150-8097}
\urldef\tempurl%
\url{https://doi.org/10.14778/3641204.3641205}
\showDOI{\tempurl}


\bibitem[\protect\citeauthoryear{Wu, Jindal, Amizadeh, Patel, Le, Qiao, and Rao}{Wu et~al\mbox{.}}{2018}]%
        {DBLP:journals/pvldb/WuJAPLQR18}
\bibfield{author}{\bibinfo{person}{Chenggang Wu}, \bibinfo{person}{Alekh Jindal}, \bibinfo{person}{Saeed Amizadeh}, \bibinfo{person}{Hiren Patel}, \bibinfo{person}{Wangchao Le}, \bibinfo{person}{Shi Qiao}, {and} \bibinfo{person}{Sriram Rao}.} \bibinfo{year}{2018}\natexlab{}.
\newblock \showarticletitle{Towards a Learning Optimizer for Shared Clouds}.
\newblock \bibinfo{journal}{\emph{Proc. {VLDB} Endow.}} \bibinfo{volume}{12}, \bibinfo{number}{3} (\bibinfo{year}{2018}), \bibinfo{pages}{210--222}.
\newblock


\bibitem[\protect\citeauthoryear{Wu and Cong}{Wu and Cong}{2021}]%
        {DBLP:conf/sigmod/WuC21}
\bibfield{author}{\bibinfo{person}{Peizhi Wu} {and} \bibinfo{person}{Gao Cong}.} \bibinfo{year}{2021}\natexlab{}.
\newblock \showarticletitle{A Unified Deep Model of Learning from both Data and Queries for Cardinality Estimation}. In \bibinfo{booktitle}{\emph{SIGMOD}}. \bibinfo{pages}{2009--2022}.
\newblock


\bibitem[\protect\citeauthoryear{Wu, Negi, Alizadeh, Kraska, and Madden}{Wu et~al\mbox{.}}{2023}]%
        {DBLP:journals/pacmmod/WuNAKM23}
\bibfield{author}{\bibinfo{person}{Ziniu Wu}, \bibinfo{person}{Parimarjan Negi}, \bibinfo{person}{Mohammad Alizadeh}, \bibinfo{person}{Tim Kraska}, {and} \bibinfo{person}{Samuel Madden}.} \bibinfo{year}{2023}\natexlab{}.
\newblock \showarticletitle{FactorJoin: {A} New Cardinality Estimation Framework for Join Queries}.
\newblock \bibinfo{journal}{\emph{Proc. {ACM} Manag. Data}} \bibinfo{volume}{1}, \bibinfo{number}{1} (\bibinfo{year}{2023}), \bibinfo{pages}{41:1--41:27}.
\newblock


\bibitem[\protect\citeauthoryear{Wu and Shaikhha}{Wu and Shaikhha}{2020}]%
        {DBLP:journals/corr/abs-2012-14743}
\bibfield{author}{\bibinfo{person}{Ziniu Wu} {and} \bibinfo{person}{Amir Shaikhha}.} \bibinfo{year}{2020}\natexlab{}.
\newblock \showarticletitle{BayesCard: {A} Unified Bayesian Framework for Cardinality Estimation}.
\newblock \bibinfo{journal}{\emph{CoRR}}  \bibinfo{volume}{abs/2012.14743} (\bibinfo{year}{2020}).
\newblock
\urldef\tempurl%
\url{https://arxiv.org/abs/2012.14743}
\showURL{%
\tempurl}


\bibitem[\protect\citeauthoryear{Wu, Yu, Yang, Zhu, Han, Li, Lian, Zeng, and Zhou}{Wu et~al\mbox{.}}{2022}]%
        {DBLP:conf/cidr/WuYYZHLLZZ22}
\bibfield{author}{\bibinfo{person}{Ziniu Wu}, \bibinfo{person}{Pei Yu}, \bibinfo{person}{Peilun Yang}, \bibinfo{person}{Rong Zhu}, \bibinfo{person}{Yuxing Han}, \bibinfo{person}{Yaliang Li}, \bibinfo{person}{Defu Lian}, \bibinfo{person}{Kai Zeng}, {and} \bibinfo{person}{Jingren Zhou}.} \bibinfo{year}{2022}\natexlab{}.
\newblock \showarticletitle{A Unified Transferable Model for ML-Enhanced {DBMS}}. In \bibinfo{booktitle}{\emph{CIDR}}.
\newblock


\bibitem[\protect\citeauthoryear{Wu, Zhu, Pfadler, Han, Li, Qian, Zeng, and Zhou}{Wu et~al\mbox{.}}{2020}]%
        {DBLP:journals/corr/abs-2011-09020}
\bibfield{author}{\bibinfo{person}{Ziniu Wu}, \bibinfo{person}{Rong Zhu}, \bibinfo{person}{Andreas Pfadler}, \bibinfo{person}{Yuxing Han}, \bibinfo{person}{Jiangneng Li}, \bibinfo{person}{Zhengping Qian}, \bibinfo{person}{Kai Zeng}, {and} \bibinfo{person}{Jingren Zhou}.} \bibinfo{year}{2020}\natexlab{}.
\newblock \showarticletitle{{FSPN:} {A} New Class of Probabilistic Graphical Model}.
\newblock \bibinfo{journal}{\emph{CoRR}}  \bibinfo{volume}{abs/2011.09020} (\bibinfo{year}{2020}).
\newblock
\urldef\tempurl%
\url{https://arxiv.org/abs/2011.09020}
\showURL{%
\tempurl}


\bibitem[\protect\citeauthoryear{Yang, Kamsetty, Luan, Liang, Duan, Chen, and Stoica}{Yang et~al\mbox{.}}{2020}]%
        {DBLP:journals/pvldb/YangKLLDCS20}
\bibfield{author}{\bibinfo{person}{Zongheng Yang}, \bibinfo{person}{Amog Kamsetty}, \bibinfo{person}{Sifei Luan}, \bibinfo{person}{Eric Liang}, \bibinfo{person}{Yan Duan}, \bibinfo{person}{Xi Chen}, {and} \bibinfo{person}{Ion Stoica}.} \bibinfo{year}{2020}\natexlab{}.
\newblock \showarticletitle{NeuroCard: One Cardinality Estimator for All Tables}.
\newblock \bibinfo{journal}{\emph{Proc. {VLDB} Endow.}} \bibinfo{volume}{14}, \bibinfo{number}{1} (\bibinfo{year}{2020}), \bibinfo{pages}{61--73}.
\newblock


\bibitem[\protect\citeauthoryear{Yang, Liang, Kamsetty, Wu, Duan, Chen, Abbeel, Hellerstein, Krishnan, and Stoica}{Yang et~al\mbox{.}}{2019}]%
        {DBLP:journals/pvldb/YangLKWDCAHKS19}
\bibfield{author}{\bibinfo{person}{Zongheng Yang}, \bibinfo{person}{Eric Liang}, \bibinfo{person}{Amog Kamsetty}, \bibinfo{person}{Chenggang Wu}, \bibinfo{person}{Yan Duan}, \bibinfo{person}{Xi Chen}, \bibinfo{person}{Pieter Abbeel}, \bibinfo{person}{Joseph~M. Hellerstein}, \bibinfo{person}{Sanjay Krishnan}, {and} \bibinfo{person}{Ion Stoica}.} \bibinfo{year}{2019}\natexlab{}.
\newblock \showarticletitle{Deep Unsupervised Cardinality Estimation}.
\newblock \bibinfo{journal}{\emph{Proc. {VLDB} Endow.}} \bibinfo{volume}{13}, \bibinfo{number}{3} (\bibinfo{year}{2019}), \bibinfo{pages}{279--292}.
\newblock


\bibitem[\protect\citeauthoryear{Ying, Cai, Luo, Zheng, Ke, He, Shen, and Liu}{Ying et~al\mbox{.}}{2021}]%
        {graphormer}
\bibfield{author}{\bibinfo{person}{Chengxuan Ying}, \bibinfo{person}{Tianle Cai}, \bibinfo{person}{Shengjie Luo}, \bibinfo{person}{Shuxin Zheng}, \bibinfo{person}{Guolin Ke}, \bibinfo{person}{Di He}, \bibinfo{person}{Yanming Shen}, {and} \bibinfo{person}{Tie{-}Yan Liu}.} \bibinfo{year}{2021}\natexlab{}.
\newblock \showarticletitle{Do Transformers Really Perform Badly for Graph Representation?}. In \bibinfo{booktitle}{\emph{NeurIPS}}. \bibinfo{pages}{28877--28888}.
\newblock


\bibitem[\protect\citeauthoryear{Zhao, Yu, He, Li, and Zhang}{Zhao et~al\mbox{.}}{2022}]%
        {DBLP:conf/sigmod/ZhaoYHLZ22}
\bibfield{author}{\bibinfo{person}{Kangfei Zhao}, \bibinfo{person}{Jeffrey~Xu Yu}, \bibinfo{person}{Zongyan He}, \bibinfo{person}{Rui Li}, {and} \bibinfo{person}{Hao Zhang}.} \bibinfo{year}{2022}\natexlab{}.
\newblock \showarticletitle{Lightweight and Accurate Cardinality Estimation by Neural Network Gaussian Process}. In \bibinfo{booktitle}{\emph{SIGMOD}}. \bibinfo{pages}{973--987}.
\newblock


\bibitem[\protect\citeauthoryear{Zhao, Christensen, Li, Hu, and Yi}{Zhao et~al\mbox{.}}{2018}]%
        {zhao2018random}
\bibfield{author}{\bibinfo{person}{Zhuoyue Zhao}, \bibinfo{person}{Robert Christensen}, \bibinfo{person}{Feifei Li}, \bibinfo{person}{Xiao Hu}, {and} \bibinfo{person}{Ke Yi}.} \bibinfo{year}{2018}\natexlab{}.
\newblock \showarticletitle{Random sampling over joins revisited}. In \bibinfo{booktitle}{\emph{SIGMOD}}. \bibinfo{pages}{1525--1539}.
\newblock


\bibitem[\protect\citeauthoryear{Zhu, Weng, Wei, Wu, Peng, Wang, Ding, Lian, Zheng, and Zhou}{Zhu et~al\mbox{.}}{2024}]%
        {10.14778/3641204.3641209}
\bibfield{author}{\bibinfo{person}{Rong Zhu}, \bibinfo{person}{Lianggui Weng}, \bibinfo{person}{Wenqing Wei}, \bibinfo{person}{Di Wu}, \bibinfo{person}{Jiazhen Peng}, \bibinfo{person}{Yifan Wang}, \bibinfo{person}{Bolin Ding}, \bibinfo{person}{Defu Lian}, \bibinfo{person}{Bolong Zheng}, {and} \bibinfo{person}{Jingren Zhou}.} \bibinfo{year}{2024}\natexlab{}.
\newblock \showarticletitle{PilotScope: Steering Databases with Machine Learning Drivers}.
\newblock \bibinfo{journal}{\emph{Proc. VLDB Endow.}} \bibinfo{volume}{17}, \bibinfo{number}{5} (\bibinfo{date}{may} \bibinfo{year}{2024}), \bibinfo{pages}{980–993}.
\newblock
\showISSN{2150-8097}
\urldef\tempurl%
\url{https://doi.org/10.14778/3641204.3641209}
\showDOI{\tempurl}


\bibitem[\protect\citeauthoryear{Zhu, Wu, Han, Zeng, Pfadler, Qian, Zhou, and Cui}{Zhu et~al\mbox{.}}{2021}]%
        {DBLP:journals/pvldb/ZhuWHZPQZC21}
\bibfield{author}{\bibinfo{person}{Rong Zhu}, \bibinfo{person}{Ziniu Wu}, \bibinfo{person}{Yuxing Han}, \bibinfo{person}{Kai Zeng}, \bibinfo{person}{Andreas Pfadler}, \bibinfo{person}{Zhengping Qian}, \bibinfo{person}{Jingren Zhou}, {and} \bibinfo{person}{Bin Cui}.} \bibinfo{year}{2021}\natexlab{}.
\newblock \showarticletitle{{FLAT:} Fast, Lightweight and Accurate Method for Cardinality Estimation}.
\newblock \bibinfo{journal}{\emph{Proc. {VLDB} Endow.}} \bibinfo{volume}{14}, \bibinfo{number}{9} (\bibinfo{year}{2021}), \bibinfo{pages}{1489--1502}.
\newblock


\end{thebibliography}

\end{document}